\documentclass[11pt,letterpaper]{article}
\pdfoutput=1

\usepackage[utf8x]{inputenc}%
\usepackage{tcolorbox,fancybox}%
\usepackage{physics}
\usepackage{ntheorem}

\usepackage[english]{babel}
\usepackage{amsmath,amssymb,amsfonts}
\numberwithin{equation}{section}
\usepackage{graphicx}
\usepackage{booktabs}
\usepackage{color,xcolor}
\usepackage[numbers,sort&compress]{natbib}

\usepackage{tikz}
\usepackage[compat=1.1.0]{tikz-feynman}
\usetikzlibrary{positioning}
\usetikzlibrary{decorations.text}
\usetikzlibrary{decorations.pathmorphing}

\usetikzlibrary{calc}
\usetikzlibrary{shapes.misc}
\tikzset{
        >=latex,
    photon/.style={decorate, decoration={snake}, draw=black, thick},
    fermionnoarrow/.style={draw=black, postaction={decorate}, thick},
    scalar/.style={draw=black, postaction={decorate}, decoration={markings,mark=at position .55 with {\arrow{>}}}, thick, dashed},
    scalarnoarrow/.style={draw=black, postaction={decorate},  thick, dashed},
    fermion/.style={draw=black, postaction={decorate},decoration={markings,mark=at position .55 with {\arrow{>}}}, thick},
    gluon/.style={decorate, draw=black, decoration={coil,amplitude=4pt, segment length=5pt}, thick},
    vertex/.style={draw,shape=circle,fill=black,minimum size=3pt,inner sep=0pt},
    fillvertex/.style={draw,shape=circle,fill=black,minimum size=5pt,inner sep=0pt},
    openvertex/.style={draw,shape=circle,minimum size=5pt,inner sep=0pt},
    blob/.style={draw=red,shape=circle,fill=red,minimum size=6pt,inner sep=0pt},
    redvertex/.style={draw=red,shape=circle,fill=red,minimum size=3pt,inner sep=0pt},
    cross/.style={cross out, draw=black,thick, minimum size=5pt, inner sep=0pt, outer sep=0pt}
}

\usepackage{physics}
\usepackage{orcidlink}
\usepackage{slashed}
\usepackage{mathrsfs}
\usepackage{enumerate}
\usepackage{mdframed}
\usepackage{url}
\usepackage[makeroom]{cancel}
\usepackage{geometry}

\usepackage{pifont}

\newtheorem*{thm-non}{Theorem}

\usepackage{hyperref} 
\hypersetup{
    colorlinks=true,       
    linkcolor=red,          
    citecolor=blue,        
    filecolor=magenta,      
    urlcolor=purple           
}
\usepackage[all]{hypcap} 

\def\beqn{\begin{eqnarray}}
\def\eeqn{\end{eqnarray}}
\def\beqs{\begin{subequations}}
\def\eeqs{\end{subequations}}
\def\beq{\begin{equation}}
\def\eeq{\end{equation}}
\def\ba{\begin{array}}
\def\ea{\end{array}}

\def\non{\nonumber\\}
\def\f{\frac}
\def\hf{\frac{1}{2}}
\def\[{\left[}
\def\]{\right]}
\def\({\left(}
\def\){\right)}
\newcommand\para{\paragraph{}}

\def\dif{\partial}

\newcommand{\rep}[1]{\mathbf{#1}}
\newcommand{\repb}[1]{\mathbf{\overline{#1}}}

\def\Bc{\mathcal{B}}

\def\Dc{\mathcal{D}}

\def\Gc{\mathcal{G}}
\def\Hc{\mathcal{H}}

\def\Lc{\mathcal{L}}

\def\Oc{\mathcal{O}}

\def\Wc{\mathcal{W}}
\def\Xc{\mathcal{X}}

\def\Zc{\mathcal{Z}}

\usepackage{ulem}

\title{
 {\bf The topological non-Abelian string \\ in the extended ${\rm SU}(N) \otimes {\rm U}(1)$ gauge sector} \\
\author{\large Zhengyang Bian$^{1}$\,\orcidlink{0009-0000-6083-5518}, Ning Chen$^{2}$\,\orcidlink{0000-0002-0032-9012}, Mian Guo$^{3}$\,\orcidlink{0009-0002-7266-8467}, Saurabh K. Shukla$^4$\,\orcidlink{0000-0001-5344-9889}, Junyi Wei$^5$\,\orcidlink{0009-0006-4237-3456
}}
\date{\small \it
$^{1}\,^2\, ^3 \,^{4}\, ^5$School of Physics, Nankai University, Tianjin, 300071, China \\
}
}

\begin{document}

\maketitle
\setlength{\parskip}{0.2ex}

\begin{abstract}
\bigskip
We study the topological non-Abelian string and its classical stability in models with two Higgs fields to achieve the symmetry breaking of ${\rm SU}(N) \otimes {\rm U}(1)_X\to {\rm SU}(N-1)  \otimes {\rm U}(1)_{X^\prime}$.
The topological origin is due to the global $\widetilde {\rm U}(1)$ symmetry and its breaking in the scalar potential.
The most generic winding configurations of arbitrary integers of $(n_1\,, n_2)$ are considered.
The stability of the topological string is analyzed based on the time-dependent perturbations to the string background and numerical solutions to the coupled Helmholtz equations of the perturbed Fourier modes. 
We present the constraints on the stable regions for the $(n_1\,, 0)$ configuration (or equivalently the $(0\,, n_2)$ configuration).
For the general winding configurations of $(n_1\neq 0\,, n_2\neq 0)$, we point out the stable regions only exist in the semilocal limit of the large mixing angle of $\vartheta_N\to \frac{\pi}{2}$.
\end{abstract}

\vspace{8.5cm}
{\emph{Emails:}\\  
$^{\,1}$\url{mronebian@mail.nankai.edu.cn}\\
$^{\,2}$\url{chenning_symmetry@nankai.edu.cn}\\
$^{\,3}$\url{guomian212025@mail.nankai.edu.cn}\\
$^{\,4}$\url{saurabhks@nankai.edu.cn}\\
$^{\, 5}$ \url{weijunyi@mail.nankai.edu.cn}\\
 }

\thispagestyle{empty}  
\newpage  
 
\setcounter{page}{1}  

\vspace{1.0cm}
\tableofcontents

\vspace*{3mm}
\section{Introduction}
\label{section:intro}
%
%

\para
Topological defects arise naturally in theories with spontaneous symmetry breaking, as described by the Kibble-Zurek mechanism~\cite{Kibble:1976sj,Zurek:1985qw}.
In $3+1$ dimensions, string solutions originate from the symmetry breaking pattern of $\Gc \to \Hc$ with a non-trivial first homotopy group $\pi_1( \Gc / \Hc )\neq \emptyset$.
The canonical example is the Abrikosov-Nielsen-Olesen (ANO) string~\cite{Abrikosov:1956sx,Nielsen:1973cs} in the Abelian Higgs model.

\para
Attempts to embed Abelian strings into spontaneously broken non-Abelian theories, such as the electroweak (EW) sector of the Standard Model (SM), face a fundamental obstacle that the EW vacuum has a trivial first homotopy group of $\pi_1({\rm SU}(2)_W \otimes {\rm U}(1)_Y / {\rm U}(1)_{\rm EM})=\emptyset$.
Consequently, the resulting string solutions, known as $Z$ strings or EW strings~\cite{Vachaspati:1991dz,Hindmarsh:1991jq,Vachaspati:1992fi,Vachaspati:1992jk}, are not topologically protected. 
Their stability depends on the Weinberg angle $\vartheta_W$ and the Higgs mass, and a stable $Z$ string requires that $\vartheta_W\to \frac{\pi}{2}$ and $m_H< m_Z$, which are incompatible with the experimental measurements.

\para
Extending the Higgs sector, for instance via a two-Higgs-doublet model (2HDM), can reintroduce a global $\widetilde {\rm U}(1)$ symmetry in the EW sector~\cite{Earnshaw:1993yu,Dvali:1993sg,Dvali:1994qf,Eto:2018tnk,Eto:2021dca}.
This symmetry allows for topological $Z$ strings carrying non-integer magnetic fluxes.
However, explicit $\widetilde {\rm U}(1)$-breaking terms in the 2HDM potential typically lead to string-wall composites~\cite{Eto:2018hhg,Eto:2018tnk}, and the stability of these configurations is not guaranteed.

\para
The efforts of studying various string and possible composite structures in the EW sector with multiple Higgs fields can be generalized to the frameworks with extended gauge symmetries.
In an early study of the flavor issue in the framework of grand unified theories~\cite{Georgi:1979md}, Georgi conjectured that a gauge group beyond the minimal ${\rm SU}(5)$~\cite{Georgi:1974sy} can embed three generational SM fermions non-trivially.
A more realistic framework based on the ${\rm SU}(8)$ was studied in Refs.~\cite{Barr:2008pn,Chen:2023qxi,Chen:2024cht}, where a single SM Higgs boson can couple to nine flavors of the SM quarks/leptons, hence to generate the observed mass hierarchies and the CKM mixing pattern.
In the detailed studies of the possible symmetry breaking patterns in this framework~\cite{Chen:2024deo,Chen:2024yhb,Chen:2024wcj,Chen:2025ezv}, the strong/weak gauge symmetries are found to extend beyond the SM ones.
One possible subgroup that describes the next new physics beyond the EW scale is based on the $\Gc_{331}\equiv {\rm SU}(3)_c \otimes {\rm SU}(3)_W \otimes {\rm U}(1)_X$, which was first proposed in Refs.~\cite{Lee:1977qs,Lee:1977tx} historically.
Furthermore, the intermediate symmetry breaking stages between the GUT scale and the EW scale are achieved by multiple Higgs fields.

\para
In a preceding work~\cite{Bian:2026tco}, we were motivated to look for the embedded non-topological string solution in a $\Gc_{331}$ model with two anti-fundamental Higgs fields carrying the identical winding numbers.
Due to the unavoidable sources from the spin magnetic couplings, the stable regions therein were found to be close to the semilocal limit for the mixing angle (to be denoted as $\vartheta_N \to \f{\pi}{2}$ in the current context).
By converting to the relations between gauge couplings, we found that the embedded non-topological string was generally unstable in the framework with the extended ${\rm SU}(N) \otimes {\rm U}(1)$ Lie groups.
In this work, we turn to the topological string configurations suggested in Refs.~\cite{Earnshaw:1993yu,Dvali:1993sg,Dvali:1994qf,Eto:2018tnk,Eto:2021dca} and perform the detailed stability analysis to the corresponding string solutions.
We shall show that a more viable ranges of $0 \lesssim \vartheta_N \lesssim \f{\pi}{2}$ are likely, with the boundary conditions for the $(n_1\,,0)$ string configurations~\footnote{The equivalent results also hold for the $(0\,, n_2)$ string configurations.}, as well as the proper choices of the parameters in the Higgs potential.

\para
The rest of the paper is organized as follows.
In Sec.~\ref{section:generics}, we set up a model with the spontaneous symmetry breaking pattern of ${\rm SU}(N) \otimes {\rm U}(1)_X \to {\rm SU}(N-1) \otimes {\rm U}(1)_{X^\prime}$ achieved by two ${\rm SU}(N)$ anti-fundamental Higgs fields.
In Sec.~\ref{section:SUN_top_string}, we obtain the classical $\Zc^\prime$ string solution for the generic ${\rm SU}(N) \otimes {\rm U}(1)_X$ model.
With the numerical solutions of the $\Zc^\prime$ string profile functions, we carry out the detailed analysis of the string stability in Sec.~\ref{section:SUN_top_string_stability}, where we employ the time-dependent perturbations to both the Higgs fields and the gauge fields.
We summarize our results and comment on future directions in Sec.~\ref{section:conclusion}.

\vspace*{3mm}
\section{The general ${\rm SU}(N) \otimes {\rm U}(1)_X \to {\rm SU}(N-1) \otimes {\rm U}(1)_{X^\prime}$ symmetry breaking pattern}
\label{section:generics}

\para
In this section, we set up our conventions for the ${\rm SU}(N) \otimes {\rm U}(1)_X \to {\rm SU}(N-1) \otimes {\rm U}(1)_{X^\prime}$ symmetry breaking pattern, with the focus on the Higgs and gauge sectors.
Two Higgs fields of $\Phi_\omega \in ( \repb{N}\,, - \f{1}{N} )$ of the ${\rm SU}(N) \otimes {\rm U}(1)_X$ are assumed for the spontaneous symmetry breaking at this stage.

\subsection{The Higgs sector in the extended ${\rm SU}(N) \otimes {\rm U}(1)_X $ model}
\label{section:SUN_Higgs}

\para
The string solutions with the extended symmetry breaking pattern of ${\rm SU}(N) \otimes {\rm U}(1)_X \to {\rm SU}(N-1) \otimes {\rm U}(1)_{X^\prime}$ were recently discussed in Ref.~\cite{Kanda:2023yyz} with one ${\rm SU}(N)$ fundamental Higgs field.
For our purpose, we assume the spontaneous symmetry breaking of ${\rm SU}( N) \otimes {\rm U}(1)_X \to {\rm SU}( N-1) \otimes {\rm U}(1)_{X^\prime}$ to be achieved by two following ${\rm SU}(N)$ anti-fundamental Higgs fields and their VEVs of
\beqn
&&  \Phi_\omega  = \( \ba{c}
    \pi_{k \,, \omega }^-  \\   \f{1}{ \sqrt{2} }  ( \sigma_\omega - i \pi_{ \Zc^\prime\,, \omega }^0  ) \ea \) \,,  ~  \langle \sigma_\omega \rangle = V_\omega \,,  ~ k =1 \,, ...\,, N-1 \,, ~ \omega = (1\,, 2 ) \,.
\eeqn 
According to Eq.~\eqref{eq:Qcharge_antifund} below, the ${\rm U}(1)_{X^\prime}$ charges are $\hat X^\prime( \pi_{k \,, \omega }^-)= - \f{1}{N-1}$ and $\hat X^\prime( \sigma_{ \omega })= \hat X^\prime( \pi_{ \Zc^\prime\,, \omega }^0) = 0$.
The ratio between two Higgs VEVs and the sum of their squares are parametrized as
\beqn
&& V_1 = \bar V c_{\tilde \beta } \,, ~ V_2 = \bar V s_{ \tilde \beta } \,, ~  \bar V^2 \equiv V_1^2 + V_2^2 \,.
\eeqn
The most generic Higgs potential with two Higgs fields reads
\beqn\label{eq:SUN_twoHiggs_potential}
 V( \Phi_\omega )&=& \hf \lambda_1 \( | \Phi_1 |^2 - \hf V_1^2 \)^2 + \hf \lambda_2 \( | \Phi_2 |^2 - \hf V_2^2 \)^2 + \hf \lambda_3 \( | \Phi_1 |^2 + | \Phi_2 |^2  - \hf \bar V^2   \)^2    \non
 && + \lambda_4 \(  | \Phi_1|^2 | \Phi_2 |^2   - ( \Phi_1^\dag \Phi_2 ) ( \Phi_2^\dag \Phi_1 )  \) + \lambda_5 \Big| ( \Phi_1^\dag \Phi_2 ) - \hf V_1 V_2 \Big|^2 \,.
\eeqn

\para
Expanding the potential in Eq.~\eqref{eq:SUN_twoHiggs_potential} around the aligned vacuum gives the CP-even Higgs squared-mass matrix
\beqn
&& M_\sigma^2 =
\left( \ba{cc}
(\lambda_1+\lambda_3)V_1^2+\hf\lambda_5V_2^2
& (\lambda_3+\hf\lambda_5)V_1V_2 \\
(\lambda_3+\hf\lambda_5)V_1V_2
& (\lambda_2+\lambda_3)V_2^2+\hf\lambda_5V_1^2
\ea \right) \,.
\eeqn
The two CP-even mass eigenstates are defined by
\beqn
&& \left( \ba{c} h_1 \\ h_2 \ea \right)
=
\left( \ba{cc}
c_\alpha & s_\alpha \\
-s_\alpha & c_\alpha
\ea \right)
\cdot
\left( \ba{c} \sigma_1 \\ \sigma_2 \ea \right) \,,
\eeqn
with squared masses of
\beqn
m_{h_1,h_2}^2
&=& \hf\bar V^2
\Bigg[
\lambda_1 c_{\tilde\beta}^2
+\lambda_2 s_{\tilde\beta}^2
+\lambda_3+\hf\lambda_5
\nonumber\\
&& {}\pm
\sqrt{
\left[
\lambda_1 c_{\tilde\beta}^2
-\lambda_2 s_{\tilde\beta}^2
+\left(\lambda_3-\hf\lambda_5\right)c_{2\tilde\beta}
\right]^2
+\left(\lambda_3+\hf\lambda_5\right)^2s_{2\tilde\beta}^2
}
\Bigg]\,.
\eeqn
The physical CP-odd Higgs boson and the $N-1$ complex ${\rm U}(1)_{X^\prime}$-charged Higgs bosons have squared masses
\beqn
&& m_A^2=\hf\lambda_5\bar V^2,
\qquad
m_{H_k^\pm}^2=\hf\lambda_4\bar V^2,
\qquad k=1,\ldots,N-1 \,.
\eeqn

\para
In the limit of $\lambda_5\to0$, an additional global $\widetilde{\rm U}(1)$ symmetry~\cite{Eto:2018tnk} emerges in the scalar potential written in Eq.~\eqref{eq:SUN_twoHiggs_potential}
\beqn
&& (\Phi_1,\Phi_2) \to (e^{-i\gamma}\Phi_1,e^{i\gamma}\Phi_2)\,.
\eeqn
Its spontaneous breaking underlies the topological non-Abelian strings.
The CP-even squared-mass matrix and its eigenvalues reduce to
\beqn
M_\sigma^2\big|_{\lambda_5=0}
&=&
\left( \ba{cc}
(\lambda_1+\lambda_3)V_1^2 & \lambda_3V_1V_2 \\
\lambda_3V_1V_2 & (\lambda_2+\lambda_3)V_2^2
\ea \right),\nonumber \\
m_{h_1,h_2}^2\big|_{\lambda_5=0}
&=& \hf\bar V^2
\Bigg[
\lambda_1 c_{\tilde\beta}^2
+\lambda_2 s_{\tilde\beta}^2+\lambda_3
\nonumber\\
&& {}\pm
\sqrt{
\left(
\lambda_1 c_{\tilde\beta}^2
-\lambda_2 s_{\tilde\beta}^2
+\lambda_3 c_{2\tilde\beta}
\right)^2
+\lambda_3^2s_{2\tilde\beta}^2
}
\Bigg]\,.
\eeqn
The physical CP-odd scalar becomes the massless Goldstone boson, $m_A^2=0$, while the charged-Higgs masses remain $m_{H_k^\pm}^2=\hf\lambda_4\bar V^2$.

\subsection{The gauge sector in the extended ${\rm SU}(N) \otimes {\rm U}(1)_X $ model}
\label{section:SUN_gauge}

\para
Let us denote the ${\rm SU}( N)$ and the ${\rm U}( 1)_X$ gauge couplings as $(g_N\,, g_X)$, and their gauge fields as $(\Wc_\mu^I \,, \Xc_\mu)$ (with $I=1\,,\dots\,, N^2-1$).
We define the covariant derivatives for the ${\rm SU}( N)$ fundamental and anti-fundamental representations as
\beqs
\beqn
 D_\mu \Psi_\Box &\equiv& \( \dif_\mu \mathbb{I}_N - i g_N \Wc_\mu^I T_{ {\rm SU}(N) }^I - i g_X \Xc_\mu \mathbb{I}_N   \Xc \) \Psi_\Box \,, \label{eq:SUN_covariant_fund} \\[2mm]
  D_\mu \Psi_{ \bar \Box} &\equiv&  \( \dif_\mu \mathbb{I}_N + i g_N \Wc_\mu^I (T_{ {\rm SU}(N) }^I)^T - i g_X \Xc_\mu \mathbb{I}_N  \bar{\Xc} \)   \Psi_{ \bar \Box } \,,\label{eq:SUN_covariant_antifund}
\eeqn
\eeqs
where they carry the ${\rm U}( 1)_X$ charges of $\Xc$ and $\bar \Xc$, respectively.

\para
The gauge fields from Eq.~\eqref{eq:SUN_covariant_antifund} can be expressed in terms of the following $N\times N$ matrices 
\beqn\label{eq:SUN_connection_gauge}
&& - g_{N } \Wc^{ I}_\mu  ( T_{ {\rm SU}(N) }^{ I} )^T + g_{X} \bar \Xc  \mathbb{I}_N  \Xc_{ \mu}  \non
&=&  - \frac{g_{N} }{ \sqrt{2} } \( \ba{cc}  \Wc_\mu^{ \hat I } ( T_{ {\rm SU}(N- 1 ) }^{ \hat I} )^T  &  0_{ ( N-1 ) \times 1} \\   0_{ 1 \times (N-1) } & 0 \ea \)   -  \frac{g_{N} }{ \sqrt{2} } \( \ba{cc} 0 \cdot \mathbb{I }_{  (N-1 ) \times ( N -1) }  & ( \Wc_\mu^{k\, *} )_{ ( N-1 ) \times 1} \\  ( \Wc_\mu^{k } )_{ 1 \times (N-1) } & 0 \ea \)   \non
 &-& \frac{ g_{N } }{ \sqrt{2 N (N-1) } } {\rm diag}  \Big( (  \Wc_\mu^{ N^2 -1}  + ( N-1 ) t_{ \vartheta_N }  \Xc_{ \mu}  ) \mathbb{I}_{ ( N - 1 ) \times ( N-1) }  \,, \non 
&& - (N -1 ) ( \Wc_\mu^{ N^2 -1}  -  t_{ \vartheta_N }  \Xc_{ \mu} )  \Big)  \,,
\eeqn
with $\bar \Xc= - \f{1}{N}$.
For the ${\rm SU}(N)$ gauge fields of $\Wc_\mu^I$, we decompose them into $\Wc_\mu^I = \{ \Wc_\mu^{\hat I} \,, \Wc_\mu^{\bar I} \,, \Wc_\mu^{N^2 - 1}  \}$, where the $\Wc_\mu^{\hat I}$ (with $\hat I=1\,, \dots\,, N^2 - 2N$) represent the massless ${\rm SU}(N-1)$ gauge bosons after this symmetry breaking stage, the $\Wc_\mu^{\bar I}$ (with $\bar I=(N-1)^2\,, \dots\,, N^2 - 2$) represent the ${\rm U}(1)_{X^\prime }$-charged massive gauge bosons, and the last $\Wc_\mu^{N^2 - 1}$ contributes to the $\Zc^\prime$ string background together with the ${\rm U}(1)_X$ gauge field of $\Xc_\mu$.
Here, we define the ${\rm SU}( N)$ mixing angle as follows
\beqn\label{eq:SUN_mixing}
&&  t_{\vartheta_N} = \f{g_X}{ c_N g_N}\,, ~ \text{with} ~  c_N = \sqrt{\hf N(N-1)}  \,.
\eeqn
The massive and massless ${\rm U}(1)_{X^\prime }$-neutral gauge bosons are related to the gauge eigenstates in terms of the mixing angle $\vartheta_N$ as
\beqn
&&   \( \ba{c} \Wc_\mu^{N^2-1} \\ \Xc_\mu \ea \) = \( \ba{cc} c_{\vartheta_N} & s_{\vartheta_N} \\ -s_{\vartheta_N} & c_{\vartheta_N} \ea \) \( \ba{c} \Zc_\mu^\prime  \\ \Xc_\mu^\prime \ea \)   \,.
\eeqn
The ${\rm U}(1)_{X^\prime}$-neutral $\Zc^\prime$ gauge boson mass squared can be expressed as
\beqn\label{eq:Zp_masses}
&& m_{\Zc^\prime }^2 = \f{ 2 g_{\Zc^\prime }^2 }{N^2}  \times \(  \sum_\omega \hf V_\omega^2 \) = \f{  g_{ \Zc^\prime }^2 }{N^2} \bar V^2  \,,
\eeqn
where we defined the coupling for the massive $\Zc^\prime$ gauge boson as follows
\beqn
g_{ \Zc^\prime }^2 &\equiv& c_N^2  g_N^2 + g_X^2  \,.
\eeqn
Correspondingly, we have
\beqn\label{eq:Zp_couplings}
&& c_N g_N = g_{ \Zc^\prime } c_{ \vartheta_N} \,, ~ g_X = g_{ \Zc^\prime } s_{ \vartheta_N }  \,.
\eeqn
The ${\rm U}(1)_{ X^\prime }$-charged $(\Wc_\mu^k \,, \Wc_\mu^{ k \, *} )$ gauge boson masses squared can be expressed as
\beqn\label{eq:offW_masses}
&& m_{\Wc}^2 = g_N^2 \times \( \sum_\omega \frac{1}{4}  V_\omega^2 \) = \f{  g_{ \Zc^\prime }^2 }{ 2 N (N-1) } c_{ \vartheta_N }^2 \bar V^2  \,.
\eeqn
The ${\rm U}(1)_{X^\prime}$ charges after this stage of symmetry breaking for the fundamental, anti-fundamental and adjoint representations read
\beqs
\beqn
\hat X^\prime ( \rep{N} )&=& {\rm diag} \(  ( \frac{1}{ ( N-1) N } + \Xc ) \mathbb{I}_{ N-1} \,, - \frac{1}{ N} + \Xc \) \,,\label{eq:Qcharge_fund}\\[2mm]
\hat X^\prime ( \repb{N} )&=& {\rm diag} \(  ( - \frac{1}{ ( N-1) N } + \bar \Xc ) \mathbb{I}_{ N-1} \,,  \frac{1}{ N} + \bar \Xc \) \,,\label{eq:Qcharge_antifund}\\[2mm]
\hat X^\prime ( {\rm Adj} )&=& {\rm diag} \(  \frac{1}{ ( N-1) N }   \mathbb{I}_{ N-1} \,, - \frac{1}{ N} \) = \sqrt{ \f{2}{ (N-1) N } } \, T_{ {\rm SU}(N) }^{ N^2 -1} \,. \label{eq:Qcharge_adj}
\eeqn
\eeqs
If there is a sequential symmetry breaking of ${\rm SU}( N-1) \otimes {\rm U}(1)_{X^\prime} \to {\rm SU}( N-2) \otimes {\rm U}(1)_{X^{\prime \prime } }$, two mixing angles of $\vartheta_N$ and $\vartheta_{N-1}$ are related as
\beqn
&& \sin\vartheta_N = \sqrt{ \f{N-2}{N} } \tan \vartheta_{ N-1 } \,. 
\eeqn

\vspace*{3mm}
\section{The topological $\Zc^\prime$ string solution in the ${\rm SU}(N)\otimes {\rm U}(1)_X$ model}
\label{section:SUN_top_string}

\subsection{The $(n_1\,, n_2)$ $\Zc^\prime$ string profile and tension}

\para
In this section, we look for the topological $\Zc^\prime$ string solution in the ${\rm SU}(N)\otimes {\rm U}(1)_X$ model.
This is due to the extra global $\widetilde {\rm U}(1)$ symmetry by setting $\lambda_5=0$ in the generic Higgs potential in Eq.~\eqref{eq:SUN_twoHiggs_potential}.
The most generic $(n_1\,, n_2)$ $\Zc^\prime$ string solution with arbitrary winding numbers of $n_1 \neq n_2$ takes the similar profile functions in the two-dimensional polar coordinates of $(r\,, \varphi)$
\beqn\label{eq:SUN_top_stringAnsatz}
&&  \Phi_{  \omega }  =  \left( 
\ba{c}  
 \mathbf{ 0 }_{N-1}   \\ 
  \phi_{ {\rm ANO} \,, \omega }  \\  \ea  \right)\,,~  \phi_{ {\rm ANO} \,, \omega }  =  \frac{ V_\omega }{ \sqrt{2} } \bar f_\omega ( r ) e^{i n_\omega  \varphi }  \,, ~  \omega = ( 1\,, 2 ) \,, \non
  && \vec \Zc_{\rm ANO}^\prime = - \frac{ \bar \zeta ( r ) }{r } \vec e_\varphi  \,,
\eeqn
while all other components of gauge fields are vanishing~\footnote{Our gauge field profile is similar to $Z$ string profile in Refs.~\cite{Vachaspati:1992fi,Kanda:2022xrz}, while differs from the $Z$ string profile in Refs.~\cite{Goodband:1995he,Eto:2024xvc} by a factor $\propto \f{1}{g_{Z}}$. It turns out the profile function in Eq.~\eqref{eq:SUN_top_stringAnsatz} is convenient to obtain the boundary conditions in the 331 model, as well as in the generic ${\rm SU}(N) \otimes {\rm U}(1)$ models.}.
We denote the covariant derivatives to anti-fundamental Higgs fields under the $(n_1\,, n_2)$ $\Zc^\prime$ string background as follows
%
%
\beqn\label{eq:SUN_ZprimeCov_antifund}
 d_m \Phi_{ \omega }&=& \( \dif_m \mathbb{I}_N + i g_N \Wc_m^{N^2-1} T^{N^2-1} + i g_X \Xc_m \f{1}{N} \mathbb{I}_N \)  \Phi_{ \omega }\non
 &\supset& \Big\{ \dif_m \mathbb{I}_N + i g_{ \Zc^\prime}  \text{diag} \[  \f{ 1}{ N-1} \( \f{1}{ N  } - s_{ \vartheta_N}^2 \) \mathbb{I}_{N-1}, \ -\f{1}{N}   \]   \Zc_{{\rm ANO}\,, m}^\prime  \Big\} \non
 && \cdot    \( \ba{c}
    0 \\ \vdots \\ 0 \\ \phi_{{\rm ANO} \,, \omega }
  \ea \) \non
  %
%
%
|  d_m \Phi_{  \omega } |^2 &=& \hf V_\omega^2  \[  \( \f{d \bar{f}_\omega  }{dr} \)^2 +  \( n_\omega + \f{ g_{ \Zc^\prime }  }{N} \bar \zeta(r) \)^2    \( \f{\bar{f}_\omega }{r} \)^2  \] \,,
\eeqn
%
%
with $m=(1\,,2)$ representing the two-dimensional Cartesian coordinates.

\para
The string tension is generally given by
\beqn\label{eq:SUN_topo_tension}
\mu_{\rm ANO}&=& \int  d^2 x \, \Big\{  \frac{1}{4} ( \Wc_{ m n}^{ N^2 - 1} )^2 +   \frac{1}{4} ( \Xc_{ mn } )^2  + | d_m \Phi_{  \omega } |^2 + V(  \Phi_{  \omega })   \Big\} \,,
\eeqn
with the covariant derivatives in the $\Zc^\prime$ string background defined in Eq.~\eqref{eq:SUN_ZprimeCov_antifund}.
Explicitly, we have
\beqs
\beqn
&& \frac{1}{4} \( \Wc_{ mn }^{N^2 - 1}  \)^2 + \frac{1}{4} \( \Xc_{  mn }  \)^2 =   \hf \( - \frac{1 }{  r } \frac{d \bar \zeta (r ) }{d r} \)^2 \,, \\[2mm]
&& V( \Phi_{ \omega }  ) =   \frac{  1}{8 } \lambda_1V_1^4 \Big( \bar f_1^2 (r) - 1  \Big)^2  + \frac{  1 }{8 } \lambda_2 V_2^4 \Big( \bar f_2^2 (r)  - 1  \Big)^2 \non
&& + \f{1}{8} \lambda_3 \Big( V_1^2 \bar f_1^2 (r) + V_2^2 \bar f_2^2 (r) - \bar V^2  \Big)^2   \,.
\eeqn
\eeqs
Thus, we find the $(n_1\,, n_2)$ $\Zc^\prime$ string tension expressed as
\beqs\label{eqs:SUN_10_tension}
\beqn
\f{ \mu_{\rm ANO} }{ 2\pi \bar V^2 }&=&    \int \xi d \xi \, \rho_{\rm total} (\xi) \,, ~ \rho_{\rm total} (\xi) =  \rho_{\zeta } (\xi) +  \rho_{\bar f_\omega } (\xi) +  \rho_V (\xi) \,,  \\[2mm]
\rho_{\zeta } (\xi) &=&   \frac{  g_{ \Zc^\prime}^2 }{ 4 N^2 } \f{1}{ \xi^2 } \( \frac{d \bar \zeta ( \xi ) }{d \xi } \)^2  \,,\label{eq:SUN_10_tension01} \\[2mm]
\rho_{\bar f_\omega } (\xi) &=&  \hf  c_{ \tilde \beta }^2 \Big[  \(  \frac{d \bar f_1 ( \xi ) }{ d  \xi } \)^2 + \( n_1  + \frac{ g_{ \Zc^\prime }  }{ N }  \bar \zeta ( \xi ) \)^2 \( \frac{\bar f_1 ( \xi ) }{ \xi }  \)^2 \Big]  \non
&& + \hf  s_{ \tilde \beta }^2 \Big[  \(  \frac{d \bar f_2 ( \xi ) }{ d  \xi } \)^2 + \(  n_2 + \frac{ g_{ \Zc^\prime }  }{ N }  \bar \zeta ( \xi ) \)^2 \( \frac{\bar f_2 ( \xi ) }{ \xi }  \)^2 \Big]  \,,\label{eq:SUN_10_tension02} \\[2mm]
\rho_V (\xi)&=& \frac{  1}{4 } \beta_1 c_{\tilde \beta }^4  \Big( \bar f_1^2 ( \xi ) - 1  \Big)^2  + \frac{  1 }{4 } \beta_2   s_{\tilde \beta }^4  \Big( \bar f_2^2 ( \xi )  - 1  \Big)^2 \non
&& + \f{1}{4 } \beta_3   \( c_{ \tilde \beta }^2 \bar f_1^2 ( \xi ) + s_{ \tilde \beta }^2 \bar f_2^2 ( \xi ) - 1   \)^2     \,,\label{eq:SUN_10_tension03}
\eeqn
\eeqs
with the dimensionless coordinates of $\xi = \f{ m_{ \Zc^\prime } }{ \sqrt{2 } } r= \f{ g_{ \Zc^\prime } \bar V }{  \sqrt{2} N }  r$, and the ratios between the Higgs self-couplings and the gauge coupling of
\beqn\label{eq:SUN_betai}
&& \beta_i \equiv \frac{ N^2 \lambda_i }{  g_{ \Zc^\prime}^2 } \,.
\eeqn
It is also straightforward to recover the non-topological string tension~\cite{Bian:2026tco} by setting $n_1=n_2$ in Eqs.~\eqref{eqs:SUN_10_tension}.

\para
The classical Euler-Lagrange field equations are
\beqs\label{eqs:SUN_topo_EOMs}
\beqn
&&  \f{ d^2 \bar f_1 (\xi )}{d \xi^2} + \f{1}{ \xi } \f{ d \bar f_1 (\xi )}{d \xi }   - \( n_1 +\f{ g_{ \Zc^\prime } }{ N } \bar \zeta ( \xi ) \)^2 \f{ \bar f_1 (\xi ) }{ \xi^2 }  =   \beta_1 c_{\tilde \beta }^2  \( \bar f_1^2 ( \xi ) - 1 \) \bar f_1 ( \xi )   \non 
&& + \beta_3 \( c_{\tilde \beta}^2 \bar f_1^2 (\xi ) + s_{\tilde \beta}^2 \bar f_2^2 (\xi ) - 1 \)  \bar f_1 (\xi )   \,, \label{eq:SUN_topo_EOM01}\\[2mm]
&&  \f{ d^2 \bar f_2 ( \xi )}{d \xi^2} + \f{1}{ \xi } \f{ d \bar f_2 ( \xi )}{d \xi }   - \( n_2 +  \f{ g_{ \Zc^\prime } }{ N} \bar \zeta ( \xi ) \)^2 \f{ \bar f_2 (\xi ) }{ \xi^2 }  = \beta_2 s_{\tilde \beta }^2  \( \bar f_2^2 ( \xi ) - 1 \) \bar f_2 (\xi ) \non
&&  + \beta_3  \( c_{\tilde \beta}^2 \bar f_1^2 (\xi ) + s_{\tilde \beta}^2 \bar f_2^2 ( \xi ) - 1 \)  \bar f_2 ( \xi )  \,,\label{eq:SUN_topo_EOM02} \\[2mm]
&&    \f{ d^2 \bar \zeta(\xi) }{d \xi^2 } - \f{1}{ \xi } \f{d \bar \zeta( \xi )}{d \xi }  = \f{2 N }{ g_{ \Zc^\prime } }  \Big[  c_{\tilde \beta}^2  \( n_1 + \f{ g_{\Zc^\prime } }{N } \bar \zeta (\xi )  \) \bar f_1^2 ( \xi ) +  s_{\tilde \beta }^2 \( n_2  + \f{ g_{\Zc^\prime } }{N } \bar \zeta (\xi )  \) \bar f_2^2 ( \xi )  \Big]  \,.\label{eq:SUN_topo_EOM03}
\eeqn
\eeqs
Close to $\xi=0$, the leading behaviors of Eqs.~\eqref{eq:SUN_topo_EOM01} and \eqref{eq:SUN_topo_EOM02} are
\beqn
&& \f{ d^2 \bar f_\omega (\xi )}{d \xi^2} + \f{1}{ \xi } \f{ d \bar f_\omega (\xi )}{d \xi }   -  n_\omega^2  \f{ \bar f_\omega (\xi ) }{ \xi^2 }  \simeq 0 \,,
\eeqn
where we neglected the regular terms.
For $n_\omega \neq 0$, one finds the profile function behaves as $\bar f_\omega (\xi) \simeq A_\omega \xi^{n_\omega}$.
For $n_\omega=0$, one expands the profile function as $\bar f_\omega (\xi)= \sum_{m=0}^\infty a_\omega^{ (m)} \xi^m$.
The term of $\f{1}{ \xi } \f{ d \bar f_\omega (\xi )}{d \xi }$ produces a singular term of $\f{a_\omega^{(1)} }{\xi}$ at $\xi=0$, which is unwanted.
Thus at $\xi=0$, the boundary conditions for the profile functions depending on different winding numbers of $(n_1\,, n_2)$ are the following
\beqs\label{eqs:SUN_topo_BCs_00}
\beqn
(n_1 \neq 0 \,, n_2 \neq 0) ~&:&~ \bar f_1(\xi =0)  = \bar f_2(\xi =0)  =  \bar \zeta(\xi= 0) = 0 \,, \label{eq:SUN_topo_BCs_11}\\[2mm]
(n_1 \neq 0 \,, n_2 = 0) ~&:&~ \bar f_1(\xi =0)  = \bar \zeta(\xi= 0) = 0 \,, ~ \f{ d \bar f_2 }{d \xi } (\xi = 0 ) =0  \,,\label{eq:SUN_topo_BCs_10}\\[2mm]
(n_1 = 0 \,, n_2 \neq 0) ~&:&~ \bar f_2(\xi =0)  = \bar \zeta(\xi= 0) = 0 \,, ~ \f{ d \bar f_1 }{d \xi } (\xi = 0 ) =0 \,.\label{eq:SUN_topo_BCs_01}
\eeqn
\eeqs
At $\xi=\infty$, the boundary conditions for the profile functions are
\beqn\label{eqs:SUN_topo_BCs_Infinity}
&&  \bar f_1( \xi= \infty ) =  \bar f_2 ( \xi= \infty ) = 1 \,, ~ \bar \zeta( \xi= \infty)= - \frac{ N}{  g_{ \Zc^\prime } } ( n_1 c_{ \tilde \beta}^2  +  n_2 s_{ \tilde \beta}^2 ) \,.
\eeqn
The boundary conditions of $ \bar f_1( \xi= \infty ) =  \bar f_2 ( \xi= \infty ) = 1$ are obtained by minimizing of $\rho_V(\xi)$ in Eq.~\eqref{eq:SUN_10_tension03}, which correspond to a full symmetry breaking pattern of ${\rm SU}(N)\otimes {\rm U}(1)\to {\rm SU}(N-1)\otimes {\rm U}(1)_{X^\prime}$ outside of the string.
To find the boundary condition for the gauge field at $\xi=\infty$, one minimizes the energy density in Eq.~\eqref{eq:SUN_10_tension02} with respect to the $\bar \zeta(\xi)$ as follows
%
%
\beqn
&& \frac{ \delta \rho_{ \bar f_\omega } }{ \delta \bar \zeta(\xi) } \propto  c_{\tilde \beta}^2 \( n_1  + \frac{ g_{ \Zc^\prime }  }{ N }  \bar \zeta ( \xi )  \) \( \f{ \bar f_1(\xi) }{\xi} \)^2 + s_{\tilde \beta}^2 \( n_2 +  \frac{ g_{ \Zc^\prime }  }{ N }  \bar \zeta ( \xi )  \) \( \f{ \bar f_2(\xi) }{\xi}  \)^2 = 0 \non
&\Rightarrow& \bar \zeta ( \xi=\infty ) =  - \f{N }{ g_{ \Zc^\prime } } ( n_1 c_{ \tilde \beta}^2 + n_2 s_{ \tilde \beta}^2 )  \,,
\eeqn
together with the conditions of $ \bar f_1( \xi= \infty ) =  \bar f_2 ( \xi= \infty ) = 1$.

\begin{figure}[htb]
\centering
\includegraphics[height=5cm]{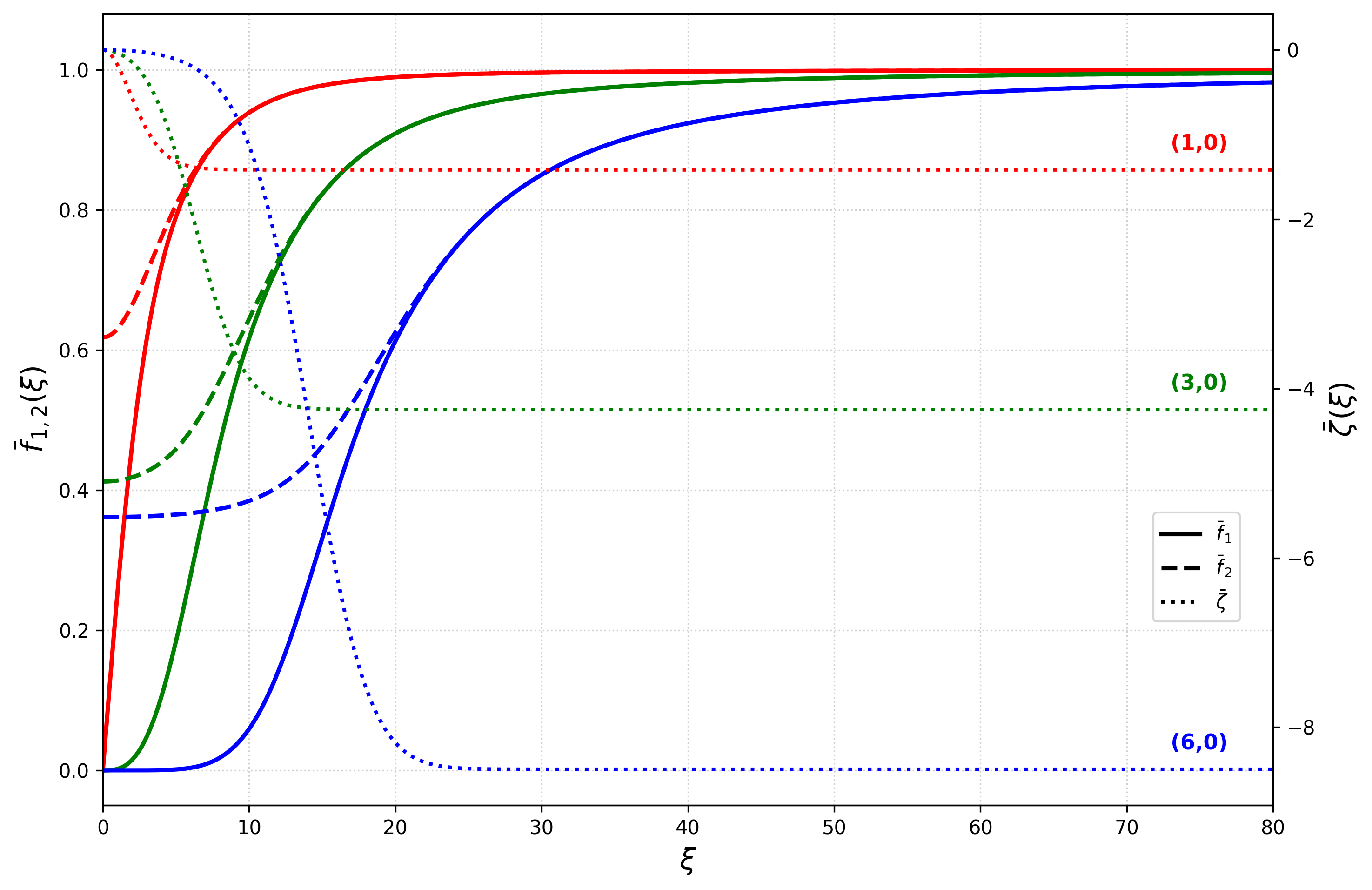}
\includegraphics[height=5cm]{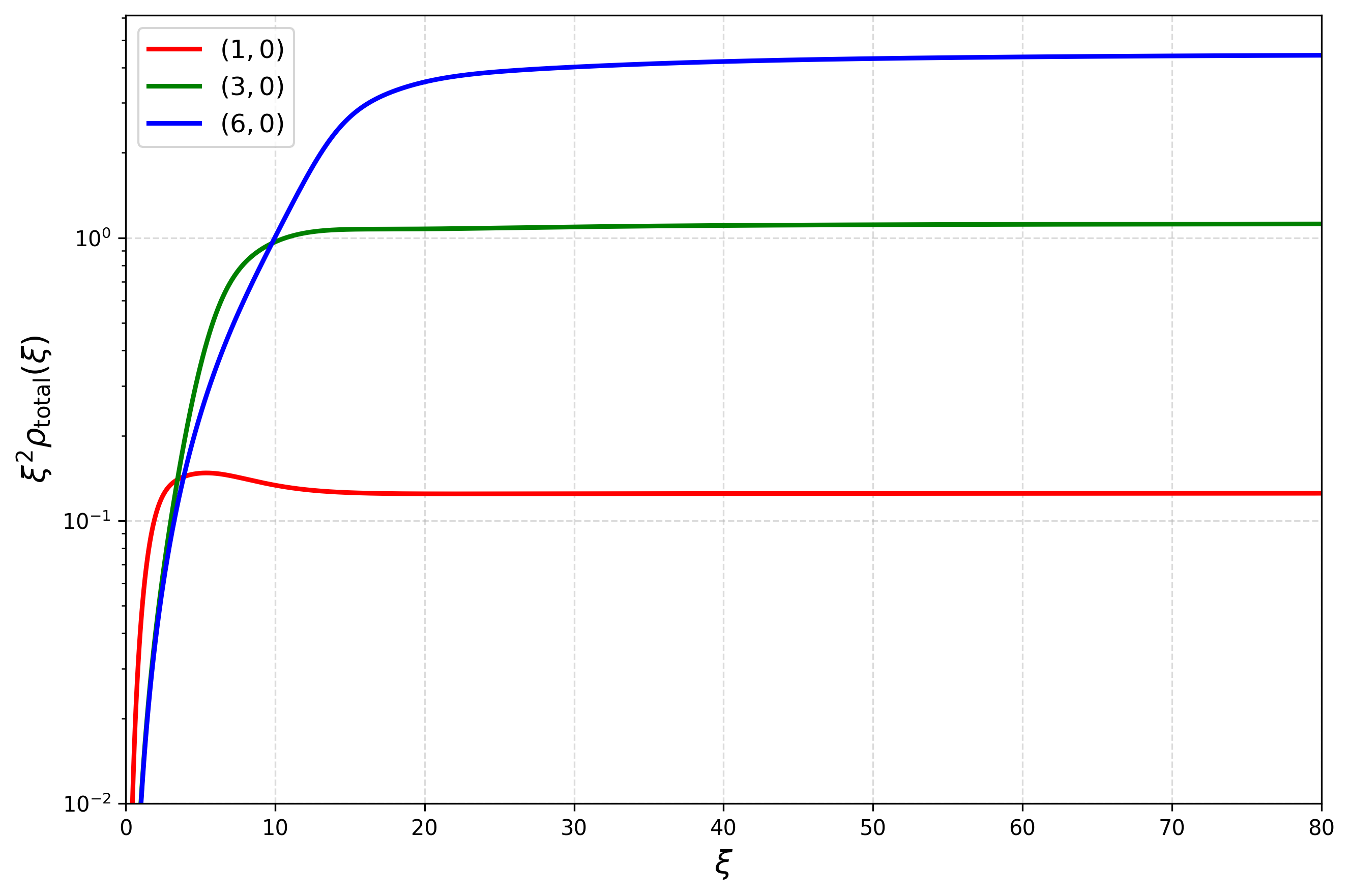}
\caption{Left panel: the string profile functions of $(\bar f_1(\xi)\,, \bar f_2 (\xi)\,, \bar \zeta(\xi) )$. 
Right panel: the total dimensionless string energy densities of $ \xi^2 \rho_{\rm total} (\xi) $.
Both plots are made for the $N=3$, and the winding configurations of $(1\,,0)$, $(3\,,0)$ and $(6\,,0)$.
Other parameters are $\lambda_1= \lambda_2 =0.15$, $\lambda_3=-0.07$, $\tilde \beta=\f{\pi}{4}$, and $g_{\Zc^\prime}=1.06$.
}
\label{fig:SUN_profiles} 
\end{figure}

\para
In the left panel of Fig.~\ref{fig:SUN_profiles}, we display the string profile functions for the winding configurations of $(1\,,0)$, $(3\,,0)$, and $(6\,,0)$, respectively.
Due to the Neumann boundary condition of Eq.~\eqref{eq:SUN_topo_BCs_10}, the profile functions of $\bar f_2(\xi)$ do not vanish at the string core of $\xi=0$.
According to the boundary conditions in Eq.~\eqref{eqs:SUN_topo_BCs_Infinity} as $\xi\to \infty$, the energy density is mainly contributed by the gauge Higgs coupling terms in Eq.~\eqref{eq:SUN_10_tension02} as follows
\beqn
&& \xi^2 \rho_{ \bar f_\omega } ( \xi ) \xrightarrow{\xi \to \infty } \hf ( n_1 - n_2)^2 c_{\tilde \beta}^2 s_{ \tilde \beta }^2 \,.
\eeqn
With $\tilde \beta= \f{\pi}{4}$ in the right panel of Fig.~\ref{fig:SUN_profiles}, the function of $\xi^2 \rho_{\rm total}(\xi)$ indeed approach to the asymptotic values of $( \f{1}{8} \,, \f{9}{8} \,, \f{9}{2})$ for the winding configurations of $(1\,,0)$, $(3\,,0)$, and $(6\,,0)$, respectively.
These finite values manifest that the topological strings are global ones with the divergent string tension~\cite{Goodband:1995rt,Eto:2018tnk}.
On the other hand, one can only expect the finite string tension with the identical winding numbers of $n_1=n_2$ as in the non-topological strings~\cite{Bian:2026tco}.

\para
For the stability analysis in Sec.~\ref{section:SUN_stable_region}, we will focus on the $(n_1\,, 0)$ string rather than the $(0\,, n_2)$ string configuration without loss of generality.
The Neumann boundary condition for the profile functions of $\bar f_2(\xi=0)$ in Eq.~\eqref{eqs:SUN_topo_BCs_00} and the Dirichlet boundary condition for the $\bar \zeta(\xi=\infty)$ in Eq.~\eqref{eqs:SUN_topo_BCs_Infinity} will play significant roles in determining the stable regions.
The more generic $(n_1\,, n_2)$ string configurations with $n_{1\,,2}\neq 0$ are usually unstable, unless the mixing angle is approaching to the semilocal limit of $\vartheta_N \to \f{\pi}{2}$.

\subsection{The non-integer magnetic flux}

\para
With the boundary condition for the $\bar \zeta( \xi = \infty)$ given in Eq.~\eqref{eqs:SUN_topo_BCs_Infinity}, the generic  $(n_1\,, n_2)$ $\Zc^\prime$ magnetic flux for the topological string is quantized as
\beqn\label{eq:SUN_magFlux}
\Phi_{ \Zc^\prime}^{( n_1\,,  n_2 )}&=& \int d^2 x \,  \partial_{ [1 } \Zc_{ {\rm ANO}\, 2] }^\prime = \int_{0 }^{ 2\pi } d\varphi \, \lim_{ r\to \infty} (r \Zc_{{\rm ANO}\,, \varphi}^\prime )  = \f{  2 \pi N  }{ g_{ \Zc^\prime}} ( n_1 c_{ \tilde \beta }^2  + n_2 s_{ \tilde \beta }^2   ) \,.
\eeqn
This is different from the integer magnetic flux of the non-topological $\Zc^\prime$ string with the identical winding numbers of $n_1=n_2$~\cite{Vachaspati:1992fi,Eto:2021dca,Bian:2026tco}, and is consistent with the non-integer magnetic flux of the EW $Z$ string in the context of the 2HDM~\cite{Dvali:1993sg}.
Specifically, the magnetic fluxes for the $( n_1\,, 0 )$ and $( 0\,, n_2)$ $\Zc^\prime$ strings read
\beqn
&& \Phi_{ \Zc^\prime}^{(  n_1\,, 0 )} = \f{  2 \pi N  }{ g_{ \Zc^\prime}}  n_1 c_{ \tilde \beta }^2  \,, ~  \Phi_{ \Zc^\prime}^{( 0 \,, n_2 )} = \f{  2 \pi N  }{ g_{ \Zc^\prime}}  n_2 s_{ \tilde \beta }^2    \,.
\eeqn
%

\vspace*{3mm}
\section{The stability analysis of the topological $\Zc^\prime$ string in the ${\rm SU}(N)\otimes {\rm U}(1)_X$ model}
\label{section:SUN_top_string_stability}

\subsection{The perturbations to the $(n_1\,, n_2)$ $\Zc^\prime$ string}
\label{section:SUN_string_perturbations}

\para
Next, we perform the perturbations to the generic $(n_1\,, n_2)$ $\Zc^\prime$ string background. 
For the Higgs fields, we expand them as
\beqn
&&    \Phi_\omega = \( \ba{c}
    \delta \pi_{k\,, \omega }^- \\
    \phi_{{\rm ANO}\,, \omega } 
  \ea \) \,,   ~ k=1\,, \dots\,, N-1 \,.
\eeqn
where we have the ${\rm U}(1)_{X^\prime }$ charges of $\hat X^\prime (\delta \pi_{k\,, \omega }^-)=- \f{1}{N-1}$ according to Eq.~\eqref{eq:Qcharge_antifund}.

\para
The ${\rm SU}(N)$ gauge fields of $\Wc_\mu^I$ follow the discussions in Sec.~\ref{section:SUN_gauge}, where the ${\rm U}(1)_{X^\prime}$-charged massive gauge bosons of $\Wc_\mu^{\bar I}$ will be treated as the perturbed fields in the gauge sector.
They will be expressed in terms of mass eigenstates as follows
\beqn
\ \( \ba{c}  \delta \Wc_\mu^{k }  \\   \delta \Wc_\mu^{k \, *} \ea \) &\equiv& \frac{1}{ \sqrt{2}}   \( \ba{cc} 1 & -i \\  1 & +i  \\ \ea \)  \cdot  \( \ba{c}   \Wc_\mu^{ \bar I = (N-1)^2 + (2 k -2 ) }  \\  \Wc_\mu^{ \bar J = (N-1)^2 + (2 k -1 ) }  \ea \) \,, ~ k = 1 \,, ... \,,  N - 1 \,, \non
&\hookrightarrow& \begin{cases} \Wc_\mu^{ \bar I = (N-1)^2 + (2 k -2 ) }  = \f{1}{\sqrt{2} } ( \delta \Wc_\mu^{k } +   \delta \Wc_\mu^{k \, *} ) \\   \Wc_\mu^{ \bar J = (N-1)^2 + (2 k -1 ) } = \f{ i }{\sqrt{2} }  ( \delta \Wc_\mu^{k } -   \delta \Wc_\mu^{k \, *} )  \end{cases}  \,.
\eeqn
Correspondingly, the field strength tensors in the gauge sector become
\beqs
\beqn
\Wc_{mn}^{\bar{I}}  &=& \dif_{ [ m }  \Wc_{n ] }^{\bar{I}} \pm \f{ g_{ \Zc^\prime } c_{\vartheta_N}^2 }{N-1}  \Wc_{[m}^{\bar{J }}   \Zc_{{\rm ANO}\,, n] }^\prime  \,, ~  \bar I = (N-1)^2\,, \dots \,, N^2 -2 \,, ~ \bar J = \bar I \pm 1  \,, \non
&\hookrightarrow&  \begin{cases}   \Wc_{mn}^k  = \f{1}{ \sqrt{2} } ( \Wc_{mn}^{ \bar I }  - i \Wc_{m n }^{ \bar I + 1 } ) = \partial_{ [m } \delta \Wc_{ n ] }^k + \f{ i g_{ \Zc^\prime} c_{ \vartheta_N }^2 }{ N -1 } \delta \Wc_{ [m }^k  \Zc_{{\rm ANO}\,, n ] }^\prime   \\   \Wc_{mn}^{k \, *}  = \f{1}{ \sqrt{2} } ( \Wc_{mn}^{ \bar I }  + i \Wc_{m n }^{ \bar I + 1 } ) = \partial_{ [m } \delta \Wc_{ n ] }^{k \, * } -  \f{ i g_{ \Zc^\prime} c_{ \vartheta_N }^2 }{ N -1 } \delta \Wc_{ [m }^{k \, * } \Zc_{{\rm ANO}\,, n ] }^\prime   \end{cases} \,,  \\[2mm]
\Wc_{mn}^{ N^2 -1 } &=& c_{\vartheta_N } \(  \dif_{ [m }  \Zc_{{\rm ANO}\,, n ]}^{ \prime }  -  \f{i g_{ \Zc^\prime } }{ N-1} \delta \Wc_{ [m }^k  \delta \Wc_{ n ]}^{k \, *}  \)   \,, 
\eeqn
\eeqs
where we used the relevant structure constants of $ f^{\bar{I} \bar{J}\,, N^2-1} = \pm \sqrt{ \frac{N }{2 (N-1) } }$ for $\bar J= \bar I \pm 1$.
The covariant derivatives to the anti-fundamental Higgs fields with the perturbed terms read
\beqn
D_m \Phi_\omega &=& \( \dif_m \mathbb{I}_N + i g_N \Wc_m^{\bar{I}} (T^{\bar{I}})^T + i g_N \Wc_m^{N^2-1} T^{N^2-1} + i g_X \Xc_m \mathbb{I}_N \f{1}{N} \) \Phi_\omega \non
%
%
&=&  \( \ba{c}
    \dif_m \delta \pi_{k\,, \omega }^-  + \f{i g_{ \Zc^\prime } c_{\vartheta_N} }{\sqrt{N(N-1)}}  \phi_{{\rm ANO}\,, \omega  }  \delta \Wc_m^{k \, *}  \\
    \underbrace{ + \f{ i g_{ \Zc^\prime } }{ N-1}  \( \f{1}{N } -  s_{\vartheta_N}^2 \) \Zc_{{\rm ANO} \,, m}^\prime \cdot \delta \pi_{k\,, \omega }^- }_{k=1, \dots, N-1 \text{ rows}} \\[2mm]
    \dif_m  \phi_{ {\rm ANO} \,, \omega }  +  \f{ i g_{ \Zc^\prime } c_{\vartheta_N} }{\sqrt{N(N-1)}} \delta \Wc_m^{k  } \cdot \delta \pi_{k \,, \omega }^- \\
    - i \f{ g_{ \Zc^\prime }}{N} \Zc_{{\rm ANO} \,, m}^\prime  \phi_{ {\rm ANO}\,, \omega } 
  \ea \) \,.
\eeqn
One can also find the ${\rm U}(1)_{X^\prime}$ charges of $\hat X^\prime (\delta \Wc_m^k  ) = + \f{1}{N-1}$ and $\hat X^\prime( \delta  \Wc_m^{k \, *}) = -  \f{1}{N-1}$ according to Eq.~\eqref{eq:Qcharge_adj}.
The expansion of the Higgs potential in Eq.~\eqref{eq:SUN_twoHiggs_potential} to the quadratic perturbed terms reads
\beqn\label{eq:SUN_twoHiggs_perturb}
V(  \delta \pi_{k \,, \omega }^\pm )&=& \hf \lambda_1 V_1^2  \( \bar f_1^2 (r )  - 1 \) \delta \pi_{k\,, 1}^+ \delta \pi_{k\,, 1}^-  +  \hf \lambda_2 V_2^2  \( \bar f_2^2 (r )  - 1 \) \delta \pi_{k\,, 2}^+ \delta \pi_{k\,, 2}^- \non
 &+& \hf \lambda_3 \( V_1^2 ( \bar f_1^2(r)- 1 ) + V_2^2 ( \bar f_2^2(r)- 1 ) \) \(  \delta \pi_{k\,,1}^- \delta \pi_{k\,,1}^+ + \delta \pi_{k\,,2}^- \delta \pi_{k\,,2}^+ \)   \non
 &+&  \f{ \lambda_4}{2}  \( V_2^2 \bar f_2^2 (r) \delta \pi_{k \,,1}^- \delta \pi_{ k\,,1}^+ + V_1^2 \bar f_1^2 (r) \delta \pi_{ k\,, 2}^- \delta \pi_{ k\,,2}^+  \right. \non
 && \left. -  V_1 V_2 \bar f_1(r) \bar f_2 (r)  (  e^{-i (n_1 - n_2) \varphi } \delta \pi_{k\,, 1}^- \delta \pi_{k \,, 2}^+ +  e^{ i (n_1 - n_2) \varphi } \delta \pi_{k \,,1}^+ \delta \pi_{ k \,,2}^- )  \)   \,.
\eeqn
Altogether, we have the string tension expanded to the quadratic perturbed terms as follows
\beqn
\mu&=& \mu_{\rm ANO} + \delta \mu_W [ \delta \Wc^k \,, \delta \Wc^{k \, * }  ] + \delta \mu_\pi [ \delta \pi_{k \,, \omega }^\pm  ] + \delta \mu_c [ \delta \Wc^k \,, \delta \Wc^{k \, * } \,,  \delta \pi_{k \,, \omega }^\pm  ] \,.
\eeqn

\para
The gauge field perturbations read
\beqn\label{eq:gauge_perturb}
%
%
\delta \mu_\Wc &=&  \int d^2 \xi \, \Big\{   \[ \partial_{[ \xi_1 } \delta  \Wc_{ \xi_2 ]}^{k } -  \f{ i g_{ \Zc^\prime } c_{\vartheta_N }^2 }{ N-1 } \f{ \bar \zeta ( \xi ) }{ \xi }  \(  \delta  \Wc_{ \xi_1 }^{ k } c_\varphi + \delta  \Wc_{ \xi_2  }^{ k } s_\varphi  \) \] \non
&& \cdot \[ \partial_{[ \xi_1 } \delta \Wc_{ \xi_2 ]}^{ k \, * } + \f{  i g_{ \Zc^\prime } c_{\vartheta_N }^2 }{ N-1 } \f{ \bar \zeta ( \xi ) }{  \xi  }  \(  \delta \Wc_{ \xi_1 }^{k \, * } c_\varphi  + \delta  \Wc_{ \xi_2 }^{ k \, * } s_\varphi \) \] \non
&&   + \f{ i g_{ \Zc^\prime } c_{\vartheta_N }^2 }{ ( N-1 ) }  \f{1}{ \xi } \f{ d \bar \zeta( \xi ) }{d \xi }  \delta \Wc_{[ \xi_1 }^{ k } \delta  \Wc_{ \xi_2 ]}^{ k \, * }    +  \f{ N c_{\vartheta_N }^2 }{ N-1 }  \( \sum_\omega \f{ V_\omega^2 }{ \bar V^2 }    \bar f_\omega^2 (\xi ) \)\delta  \Wc_{\xi_m}^{ k }  \delta W_{\xi_m}^{ k \, *}    \Big\} \,.
\eeqn

\para
The pure scalar perturbations come from the covariant derivative terms and the Higgs potential only containing the perturbed scalar components
\beqn\label{eq:scalar_perturb}
\delta \mu_\pi &=&   \sum_\omega \int  d^2 \xi  \, \Big\{   \Big| \Big( \partial_{\xi_1} + \f{ i g_{ \Zc^\prime } }{ N-1} \(  \frac{1}{N} -  s_{\vartheta_N }^2 \)  \( \f{\bar \zeta ( \xi)}{\xi } s_\varphi   \)  \Big) \delta \pi_{k \,, \omega }^-  \Big|^2 \non
 && +  \Big|  \Big( \partial_{\xi_2 } + \f{ i g_{ \Zc^\prime } }{ N-1} \(  \frac{1}{N} -  s_{\vartheta_N }^2 \) \( - \f{\bar \zeta ( \xi)}{\xi } c_\varphi   \)   \Big) \delta \pi_{ k \,, \omega }^-  \Big|^2 + \widetilde V( \delta \pi_{ k\,, \omega }^\pm  )    \Big\} \,,
\eeqn
with the perturbed Higgs potential $\sum_\omega \widetilde V( \delta \pi_{ k\,, \omega }^\pm  )$ given as
\beqn\label{eq:SUN_twoHiggs_perturb_reduced}
\sum_\omega \widetilde V(  \delta \pi_{k \,, \omega }^\pm )&=& \beta_1 c_{\tilde \beta }^2  \( \bar f_1^2 ( \xi )  - 1 \) \delta \pi_{k\,, 1}^+ \delta \pi_{k\,, 1}^-  +    \beta_2 s_{\tilde \beta }^2  \( \bar f_2^2 ( \xi )  - 1 \) \delta \pi_{k\,, 2}^+ \delta \pi_{k\,, 2}^- \non
 &+& \beta_3 \( c_{\tilde \beta }^2 ( \bar f_1^2(\xi)- 1 ) + s_{\tilde \beta }^2 ( \bar f_2^2( \xi )- 1 ) \) \(  \delta \pi_{k\,,1}^- \delta \pi_{k\,,1}^+ + \delta \pi_{k\,,2}^- \delta \pi_{k\,,2}^+ \)  \non
 &+&  \beta_4  \( s_{\tilde \beta }^2 \bar f_2^2 (\xi ) \delta \pi_{k \,,1}^- \delta \pi_{ k\,,1}^+ + c_{\tilde \beta }^2 \bar f_1^2 ( \xi ) \delta \pi_{ k\,, 2}^- \delta \pi_{ k\,,2}^+  \right. \non
 && \left. -  s_{ \tilde \beta } c_{ \tilde \beta }  \bar f_1( \xi ) \bar f_2 ( \xi )  (  \delta \pi_{k\,, 1}^- \delta \pi_{k \,, 2}^+ + \delta \pi_{k \,,1}^+ \delta \pi_{ k \,,2}^- )  \)  \,.
\eeqn

\para
The perturbed couplings between the scalars and the gauge fields read
\beqn\label{eq:scalar_gauge_perturb}
\delta \mu_c &=&  i  c_{\vartheta_N } \sqrt{ \f{ N  }{   N-1 }   } \sum_\omega  \frac{ V_\omega }{  \bar V }    \int d^2 \xi  \Big\{  \bar f_\omega ( \xi )  \( e^{i n_\omega \varphi } ( \partial_{\xi_m} \delta \pi_{k \,, \omega }^+   )  \delta \Wc_{\xi_m}^{ k\, * }  -  e^{ - i n_\omega \varphi }  ( \partial_{\xi_m } \delta \pi_{k \,, \omega }^-   )  \delta \Wc_{\xi_m}^{ k } \)   \non
&& +  \partial_{\xi_m } ( \bar f_\omega (\xi ) e^{ - i n_\omega \varphi } )   \cdot \delta \Wc_{\xi_m}^{ k } \delta \pi_{ k \,, \omega }^-  -   \partial_{\xi_m } ( \bar f_\omega ( \xi ) e^{i n_\omega \varphi } )  \cdot \delta \Wc_{\xi_m}^{ k \, * } \delta \pi_{ k \,, \omega }^+   \non
&& +  \f{ i g_{ \Zc^\prime } }{N-1} \(  \frac{ N-2 }{  N } +  s_{ \vartheta_N }^2  \)   \bar f_\omega ( \xi ) \f{ \bar \zeta (\xi ) }{\xi }   \Big[  s_\varphi   \( e^{ - i n_\omega \varphi } \delta \Wc_{\xi_1}^{k } \delta \pi_{k \,, \omega }^-  +  e^{i n_\omega \varphi }  \delta \Wc_{\xi_1}^{ k \, * } \delta \pi_{k \,, \omega }^+   \) \non
&&   -  c_\varphi  \(e^{ - i n_\omega \varphi }    \delta  \Wc_{\xi_2}^{k } \delta \pi_{k \,, \omega }^-  +  e^{i n_\omega \varphi }  \delta  \Wc_{\xi_2}^{ k \, * } \delta \pi_{k \,, \omega }^+   \)  \Big]  \Big\} \,.
\eeqn
The first line above contains the linear derivative terms between the scalars and the gauge fields, which will be removed by introducing the following gauge fixing terms
\beqn\label{eq:SUN_fix}
\Lc_{\rm fix}&=&   F( \delta \Wc_m^{k }) \cdot F( \delta \Wc_m^{k\, * })   \,, \non
%
%
F( \delta \Wc_m^{k })  &=&  \f{ m_{Z^\prime}}{ \sqrt{2} } \( \vec \nabla_\xi \cdot \delta \vec \Wc^k - \f{ i g_{ \Zc^\prime } c_{ \vartheta_N }^2 }{ N-1}  \f{\bar \zeta (\xi) }{ \xi} ( s_\varphi \delta \Wc_{\xi_1 }^k  - c_\varphi \delta \Wc_{ \xi_2 }^k  ) \right. \non 
&& \left. + i c_{\vartheta_N } \sqrt{ \f{N}{ N-1} } \sum_\omega \f{ V_\omega }{ \bar V } \bar f_\omega ( \xi ) e^{ i n_\omega \varphi } \delta \pi_{k \,, \omega }^+    \) \,, \non
%
%
F( \delta \Wc_m^{k\, * }) &=& \f{ m_{Z^\prime}}{ \sqrt{2} } \(  \vec \nabla_\xi \cdot \delta \vec \Wc^{k \, * } + \f{ i g_{ \Zc^\prime } c_{ \vartheta_N }^2 }{ N-1}  \f{\bar \zeta (\xi) }{ \xi} ( s_\varphi \delta \Wc_{\xi_1 }^{k\, *}  - c_\varphi \delta \Wc_{ \xi_2 }^{k \, * }  ) \right. \non 
&& \left. - i c_{\vartheta_N } \sqrt{ \f{N}{ N-1} } \sum_\omega \f{ V_\omega }{ \bar V } \bar f_\omega ( \xi ) e^{ - i n_\omega \varphi } \delta \pi_{k \,, \omega }^-  \)   \,.
\eeqn
Below, we transform the gauge fields to the polar coordinates by
\beqn
&& \left(  \ba{c}  \delta \Wc_{\xi_1}^{k } / \delta \Wc_{\xi_1}^{k \, *}   \\[1mm]   \delta \Wc_{\xi_2}^{ k } / \delta \Wc_{\xi_2}^{ k \, *} \ea \right) =  \left(  \ba{cc}  c_\varphi  &  - s_\varphi  \\ s_\varphi   & c_\varphi  \ea \right) \cdot \left(  \ba{c}   \delta \Wc_\xi^{k } /  \delta \Wc_\xi^{k \, *}   \\[1mm]   \delta \Wc_\varphi^{k } / \delta \Wc_\varphi^{k \, * }   \ea \right)   \,.
\eeqn

\para
The pure gauge perturbations in Eq.~\eqref{eq:gauge_perturb} are modified into
\beqn\label{eq:gauge_perturb_fixed}
\delta \tilde \mu_\Wc &=&  \int d^2 \xi \, \Big\{  \(  \f{ \partial \delta \Wc_\varphi^k }{ \partial \xi } + \f{1}{ \xi} ( \delta \Wc_\varphi^k  - \f{ \partial \delta \Wc_\xi^k }{ \partial \varphi } )  \) \( \f{ \partial \delta \Wc_\varphi^{k \, *} }{ \partial \xi } + \f{1}{ \xi } ( \delta \Wc_\varphi^{ k\, *}   - \f{ \partial \delta \Wc_\xi^{k \, *}  }{ \partial \varphi } )  \)  \non
&& +   \( \f{ \partial \delta \Wc_\xi^{k } }{\partial \xi } +  \f{ 1 }{ \xi } (\delta \Wc_\xi^{k } + \f{ \partial \delta \Wc_\varphi^{k } }{ \partial \varphi } ) \) \cdot \( \f{ \partial \delta \Wc_\xi^{k \, *} }{\partial \xi } +  \f{ 1 }{ \xi  }  (\delta \Wc_\xi^{k \, * } + \f{ \partial \delta \Wc_\varphi^{k \, * } }{ \partial \varphi } ) \)  \non
&& +  \Big[ \frac{ g_{ \Zc^\prime }^2  c_{\vartheta_N }^4 }{  ( N-1)^2 }   \( - \f{ \bar \zeta ( \xi ) }{ \xi }  \)^2 +  \frac{  N c_{\vartheta_N }^2  }{   N-1 }  \( \sum_\omega \f{ V_\omega^2 }{ \bar V^2 }    \bar f_\omega^2 ( \xi )  \) \Big] \(  | \delta \Wc_\xi^k |^2   +  |  \delta \Wc_\varphi^{k } |^2  \)   \non
 && - \frac{i g_{ \Zc^\prime }  c_{\vartheta_N }^2 }{ ( N-1 ) }  \Big[ \f{\bar \zeta ( \xi ) }{ \xi }    \delta \Wc_\xi^k    \( \f{ \partial \delta \Wc_\varphi^{k \, *} }{ \partial \xi  } + \f{1}{ \xi } ( \delta \Wc_\varphi^{ k\, *}   -  \f{ \partial \delta \Wc_\xi^{k \, *}  }{ \partial \varphi } )  \)  -  \f{\bar \zeta ( \xi ) }{ \xi }  \delta \Wc_\xi^{k\, * } \( \f{ \partial \delta \Wc_\varphi^{k } }{ \partial \xi } + \f{1}{ \xi } ( \delta \Wc_\varphi^{ k }   - \f{ \partial \delta \Wc_\xi^{k }  }{ \partial \varphi } )  \)   \non
 &&  -  \( \f{1}{ \xi } \f{d \bar \zeta ( \xi ) }{ d \xi } \) \cdot  \( \delta \Wc_{\xi }^{k } \delta \Wc_{\varphi }^{k \, * }  -  \delta \Wc_{ \varphi }^{k } \delta \Wc_{ \xi }^{k \, * }  \)   \Big]  \non
&& +  \f{ i g_{ \Zc^\prime } c_{ \vartheta_N }^2 }{ N-1} \( - \f{ \bar \zeta (\xi ) }{ \xi } \) \cdot  \Big[  \( \f{ \partial \delta \Wc_\xi^{k } }{\partial \xi } +  \f{ 1 }{ \xi } (\delta \Wc_\xi^{k } + \f{ \partial \delta \Wc_\varphi^{k } }{ \partial \varphi } ) \) \delta \Wc_\varphi^{k \, *} \non
&&  - \( \f{ \partial \delta \Wc_\xi^{k \, *} }{\partial \xi } +  \f{ 1 }{ \xi  }  (\delta \Wc_\xi^{k \, * } + \f{ \partial \delta \Wc_\varphi^{k \, * } }{ \partial \varphi } )  \)   \delta \Wc_\varphi^{k } \Big]  \Big\}  \,.
\eeqn

\para
The pure scalar perturbations in Eq.~\eqref{eq:scalar_perturb} become the following
\beqn\label{eq:scalar_perturb_fixed}
%
\delta \tilde \mu_\pi  &=& \int d^2 \xi \, \Big\{ \sum_\omega \delta \pi_{k \,, \omega }^- \cdot \(   - \f{ \partial^2 }{ \partial \xi^2} - \f{1}{ \xi } \f{\partial }{ \partial \xi }  - \f{1}{ \xi^2} \f{ \partial^2 }{ \partial \varphi^2 } \right. \non
&& \left. - \f{2 i g_{\Zc^\prime } }{ N-1 } \( \f{1}{N} - s_{\vartheta_N }^2 \) \f{ \bar \zeta ( \xi ) }{ \xi^2 } \f{ \partial }{ \partial \varphi } + \f{ g_{ \Zc^\prime }^2 }{ ( N-1)^2 } \( \f{1}{N} - s_{\vartheta_N }^2 \)^2  \( - \f{ \bar \zeta ( \xi ) }{ \xi } \)^2   \)   \delta \pi_{k \,, \omega }^+  \non
&& + \f{ N c_{\vartheta_N }^2 }{ N-1} \( \sum_{\omega_1} \f{V_{\omega_1} }{ \bar V } \bar f_{\omega_1}  ( \xi ) e^{ i n_{\omega_1} \varphi } \delta \pi_{k \,, \omega_1 }^+  \)  \( \sum_{\omega_2} \f{ V_{\omega_2} }{\bar V } \bar f_{\omega_2} ( \xi ) e^{ - i n_{\omega_2 } \varphi } \delta \pi_{k \,, \omega_2  }^- \)  \non
&&  + \sum_\omega \widetilde V(  \delta \pi_{k \,, \omega }^\pm )  \Big\}\,,
\eeqn
where the perturbed Higgs potential in the dimensionless coordinates of $\sum_\omega \widetilde V(  \delta \pi_{k \,, \omega }^\pm ) $ was previously given in Eq.~\eqref{eq:SUN_twoHiggs_perturb_reduced}.

\para
The perturbed couplings between scalars and the gauge fields in Eq.~\eqref{eq:scalar_gauge_perturb} are modified into
\beqn\label{eq:scalar_gauge_perturb_fixed}
\delta \tilde \mu_c &=&  2 c_{\vartheta_N }  \sqrt{ \f{N}{N-1} } \sum_\omega  \f{ V_\omega }{\bar V}  \int d^2 \xi \,    \Big[   \( n_\omega  +  \f{   g_{ \Zc^\prime }   }{ N }   \bar \zeta ( \xi ) \)  \f{ \bar f_\omega ( \xi ) }{ \xi }   \cdot   \(   e^{ - i n_\omega \varphi }  \delta \Wc_\varphi^{k }  \delta \pi_{k \,, \omega }^-  + e^{i n_\omega \varphi }  \delta \Wc_\varphi^{k \, *}   \delta \pi_{k \,, \omega }^+     \)  \non
&& + i \f{d \bar f_\omega ( \xi ) }{d \xi } \( e^{ -i n_\omega \varphi }   \delta \Wc_\xi^k   \delta \pi_{k \,, \omega }^-  -  e^{  i n_\omega \varphi } \delta \Wc_\xi^{k \, *}  \delta \pi_{k \,, \omega }^+ \)  \Big]   \,.
\eeqn

\subsection{The stability matrix}
\label{section:SUN_stablematrix}

\para
For the stability analysis, it is convenient to adopt the spin eigenstates of the perturbed gauge fields of
\beqn\label{eq:SUN_spinstates}
&& \delta \Wc_{ \uparrow }^{k }= \f{ e^{-i \varphi} }{ \sqrt{2} } ( \delta \Wc_\xi^{ k } - i \delta \Wc_\varphi^{ k } ) \,, ~ \delta \Wc_{ \downarrow }^{k }= \f{ e^{ i \varphi} }{ \sqrt{2} } ( \delta \Wc_\xi^{ k } + i \delta \Wc_\varphi^{ k } )  \,, \non
&& \delta \Wc_{ \uparrow }^{k\, * }= \f{ e^{-i \varphi} }{ \sqrt{2} } ( \delta \Wc_\xi^{ k \, *} - i \delta \Wc_\varphi^{ k \, * } ) \,, ~ \delta \Wc_{ \downarrow }^{k\, * }= \f{ e^{ i \varphi} }{ \sqrt{2} } ( \delta \Wc_\xi^{ k \, * } + i \delta \Wc_\varphi^{ k \, * } )  \,,
\eeqn
such that $( \delta \Wc_{ \uparrow }^k )^\dag =  \delta \Wc_{ \downarrow }^{k \, * }$ and $( \delta \Wc_{ \downarrow }^k )^\dag =  \delta \Wc_{ \uparrow }^{k \, * }$.

\para
We start to express the Fourier expansions to the scalars and gauge fields with independent modes as follows
\beqn\label{eq:SUN_FourierModes}
&& \delta \pi^{+}_{k \,, \omega}=\sum_{ \ell } s_{\omega \,, \ell } ( \xi ) e^{ - i ( \ell + n_\omega  )\varphi} \,, ~ \delta \pi^{-}_{k \,, \omega}= \sum_{\ell } s^*_{ \omega \,, \ell } ( \xi ) e^{  i (\ell + n_\omega ) \varphi}\,, ~ \omega = (1\,,2) \,, \non
&&   \delta \Wc^{k}_{\uparrow}=\sum_m -iw_{\uparrow \,, m } ( \xi ) e^{  i( m - 1 )\varphi} \,, ~ \delta \Wc^{k}_\downarrow=\sum_m  iw_{\downarrow \,, - m } ( \xi ) e^{  i( m + 1)\varphi} \,, \non
&& \delta \Wc^{k\, *}_\uparrow=\sum_m  -i w^*_{\downarrow \,, - m } ( \xi ) e^{ - i ( m + 1 )\varphi} \,, ~ \delta \Wc^{k\, *}_\downarrow=\sum_m  iw^*_{\uparrow \,,  m  } ( \xi ) e^{ - i( m  - 1)\varphi}  \,.
\eeqn
From the couplings in Eq.~\eqref{eq:scalar_gauge_perturb_fixed}, we find all terms are proportional to
\beqn
&&  \int d\varphi \, \sum_{ m \,, \ell } e^{ -i n_\omega \varphi } \( e^{i \varphi} ( -i w_{ \uparrow\,, m} ) e^{ i ( m -1) \varphi } - e^{-i \varphi }   ( i w_{ \downarrow\,, - m} ) e^{ i ( m + 1) \varphi }   \) e^{i ( \ell + n_\omega  ) \varphi }  \propto \delta_{\ell, -m} \,.
\eeqn
It means the couplings in Eq.~\eqref{eq:scalar_gauge_perturb_fixed} enforce the relation of $m=-\ell$ for the Fourier modes of the gauge fields as follows
\beqn\label{eq:SUN_Gauge_FourierModes}
&&   \delta \Wc^{k}_{\uparrow}=\sum_\ell -iw_{\uparrow \,, -\ell } ( \xi )  e^{ - i(  \ell + 1 )\varphi} \,,  ~ \delta \Wc^{k}_\downarrow=\sum_\ell  iw_{\downarrow \,,  \ell } ( \xi )  e^{  - i(  \ell - 1)\varphi} \,, \non
&& \delta \Wc^{k\, *}_\uparrow=\sum_\ell  -iw^*_{\downarrow \,,  \ell } ( \xi )  e^{  i (  \ell - 1 ) \varphi} \,,~ \delta \Wc^{k\, *}_\downarrow=\sum_\ell  iw^*_{\uparrow \,, - \ell  } ( \xi )  e^{  i(  \ell + 1)\varphi}  \,.
\eeqn
Conversely, we have
\beqn
&& \delta \Wc_\xi^k   = \f{ i }{ \sqrt{2} } \sum_\ell  \( - w_{ \uparrow\,, - \ell } ( \xi )  +  w_{\downarrow\,, \ell } ( \xi )   \) e^{ -i \ell \varphi } \,, ~  \delta \Wc_\varphi^k   = \f{1}{ \sqrt{2} } \sum_\ell \(  w_{ \uparrow\,, - \ell } ( \xi )  +  w_{\downarrow\,, \ell } ( \xi )   \)  e^{ -i \ell \varphi } \,, \non
%
%
&& \delta \Wc_\xi^{k \, * }  = \f{ i }{ \sqrt{2} } \sum_\ell  \( - w_{ \downarrow\,,  \ell }^* ( \xi )   + w_{\uparrow\,, - \ell }^*  ( \xi )  \)  e^{  i \ell \varphi } \,, ~ \delta \Wc_\varphi^{k \, *}  = \f{1}{ \sqrt{2} } \sum_\ell  \(   w_{ \downarrow\,,  \ell }^* ( \xi )   + w_{\uparrow\,, - \ell }^* ( \xi )     \)  e^{  i \ell \varphi }  \,.
%
\eeqn
For consistency, one can set $n_1=n_2=1$ in Eq.~\eqref{eq:SUN_FourierModes}, and shift $\ell\to \ell-1$ in both \eqref{eq:SUN_FourierModes} and \eqref{eq:SUN_Gauge_FourierModes}.
They recover the Fourier expansions that were previously given in Ref.~\cite{Bian:2026tco} for the $(1\,,1)$ winding configuration.

\para
In the basis of $\delta \Theta \equiv ( s_{1\,, \ell} \,, s_{2\,, \ell} \,, w_{ \uparrow \,, - \ell} \,, w_{\downarrow\,, \ell} )^T$ and $\delta \Theta^\dag \equiv ( s_{1\,, \ell}^* \,, s_{2\,, \ell}^* \,, w_{\uparrow\,, - \ell }^* \,, w_{\downarrow\,,  \ell }^*)$, the perturbed string tension is expressed in terms of the stability matrix as follows
\beqn
\delta \tilde \mu&=&  \delta \tilde \mu_\Wc + \delta \tilde \mu_\pi + \delta \tilde \mu_c = 2\pi (N-1) \int \xi d \xi \, \delta \Theta^\dag \hat \Oc \delta \Theta \,, \non
\hat \Oc &=& \begin{pmatrix}
\mathcal{D}_{11} & \Dc_{12} & \Bc_{ \uparrow\,, 1} &  \Bc_{ \downarrow\,, 1} \\
\Dc_{12} &  \Dc_{22} & \Bc_{ \uparrow\,, 2}   &  \Bc_{ \downarrow \,, 2} \\
\Bc_{ \uparrow \,, 1}  & \Bc_{ \uparrow \,, 2} & \mathcal{D}_{ \uparrow} & 0\\
 \Bc_{ \downarrow \,, 1} & \Bc_{ \downarrow\,, 2 }  &0  & \mathcal{D}_{ \downarrow } 
\end{pmatrix}   \,,  
\eeqn
where each matrix element reads
\beqs\label{eqs:SUN_matrix_stable}
\beqn
\mathcal{D}_{11}&=& - \frac{ d^2 }{ d \xi^2} - \frac{1}{\xi}\frac{ d }{ d \xi } +\frac{1}{\xi^2} \( n_1 + \ell+   \f{ g_{ \Zc^\prime} }{ N-1} (-\frac{1}{ N } +  s_{\vartheta_N}^2)\bar{\zeta} ( \xi )  \)^2 +\frac{N c_{\vartheta_N }^2 }{ N-1 }  c_{\tilde \beta }^2 \bar{f}_{1}^2 ( \xi )  \non
&& + \beta_1 c_{\tilde \beta}^2 ( \bar f_1^2 ( \xi) -1) + \beta_3 \(   c_{\tilde \beta}^2 ( \bar f_1^2 ( \xi) -1) +  s_{\tilde \beta}^2 ( \bar f_2^2 ( \xi) -1) \) + \beta_4 s_{\tilde \beta}^2 \bar f_2^2 ( \xi )  \,, \label{eq:SUN_stable_D11}\\[2mm]
\mathcal{D}_{22}&=& - \frac{ d^2}{ d \xi^2} - \frac{1}{\xi}\frac{ d }{ d \xi} +\frac{1}{\xi^2} \( n_2 + \ell+   \f{ g_{ \Zc^\prime} }{ N-1} (-\frac{1}{ N } +  s_{\vartheta_N}^2)\bar{\zeta} ( \xi )  \)^2 +\frac{N c_{\vartheta_N }^2 }{ N-1 }  s_{\tilde \beta }^2 \bar{f}_{2}^2 ( \xi )  \non
&& + \beta_2 s_{\tilde \beta}^2 ( \bar f_2^2 ( \xi) -1) +   \beta_3 \(   c_{\tilde \beta}^2 ( \bar f_1^2 ( \xi) -1) +  s_{\tilde \beta}^2 ( \bar f_2^2 ( \xi) -1) \) + \beta_4 c_{\tilde \beta}^2 \bar f_1^2 ( \xi ) \,,\label{eq:SUN_stable_D22} \\[2mm]
\mathcal{D}_{12}&=& \[  - \beta_4    \bar f_1( \xi ) \bar f_2 ( \xi )  + \f{ N c_{\vartheta_N }^2 }{ N-1  }  \bar f_1( \xi ) \bar f_2 ( \xi ) \]  s_{\tilde \beta} c_{\tilde \beta}  \,,\label{eq:SUN_stable_D12} \\[2mm]
\Bc_{ \uparrow\,, 1}&=& \sqrt{ \f{2 N}{ N-1 } } c_{\vartheta_N } c_{\tilde \beta} \left[  \frac{ d \bar{f}_1 ( \xi ) }{ d \xi}+\frac{ \bar{f}_1( \xi ) }{\xi} \( n_1 +\frac{g_{ \Zc^\prime }}{ N } \bar{\zeta} ( \xi ) \) \right] \,, \\[2mm]
\Bc_{ \downarrow\,, 1 }&=& \sqrt{ \f{2 N}{ N-1 } }  c_{\vartheta_N } c_{\tilde \beta}  \left[ -\frac{d  \bar{f}_1 (\xi) }{ d \xi}+\frac{ \bar{f}_1( \xi ) }{\xi} \( n_1 +\frac{g_{ \Zc^\prime }}{ N  }\bar{\zeta} (\xi) \) \right] \,, \\[2mm]
\Bc_{ \uparrow\,, 2}&=& \sqrt{ \f{2 N}{ N-1 } } c_{\vartheta_N } s_{\tilde \beta} \left[  \frac{ d \bar{f}_2 (\xi) }{d \xi}+\frac{ \bar{f}_2 ( \xi ) }{\xi} \( n_2 +\frac{g_{ \Zc^\prime }}{ N } \bar{\zeta} ( \xi ) \) \right] \,, \\[2mm]
\Bc_{ \downarrow\,, 2 }&=& \sqrt{ \f{2 N}{ N-1 } }  c_{\vartheta_N } s_{\tilde \beta} \left[ -\frac{d \bar{f}_2 (\xi) }{d \xi}+\frac{ \bar{f}_2 ( \xi ) }{\xi} \( n_2 +\frac{g_{ \Zc^\prime }}{ N  }\bar{\zeta} (\xi) \) \right] \,, \\[2mm]
\mathcal{D}_{ \uparrow } &=& -\frac{d^2}{d \xi^2}- \frac{1}{\xi}\frac{ d }{ d \xi}+\frac{1}{\xi^2} \( \ell +1 - \frac{ g_{ \Zc^\prime} c_{\vartheta_N}^2 }{ N-1 }  \bar{\zeta} ( \xi) \)^2  \non 
&& + \f{ 2 g_{ \Zc^\prime } c_{\vartheta_N}^2 }{ N-1 } \frac{1}{\xi}\frac{ d \bar{\zeta} ( \xi ) }{d \xi}+\frac{ N c_{\vartheta_N }^2 }{ N-1  } \sum_\omega  \bar{f}_{\omega}^2 ( \xi ) \frac{ V^2_\omega}{ \bar V^2}\,,\label{eq:SUN_matrix_Dup} \\[2mm]
\mathcal{D}_{ \downarrow }&=& - \frac{ d^2}{ d \xi^2} - \frac{1}{\xi}\frac{ d }{ d \xi}+\frac{1}{\xi^2} \( \ell - 1 - \frac{ g_{ \Zc^\prime }  }{ N-1 } c_{\vartheta_N}^2  \bar{\zeta} ( \xi ) \)^2 \non
&&  - \f{2 g_{Z^\prime } c_{\vartheta_N }^2 }{N-1} \frac{1}{\xi}\frac{ d \bar{\zeta} ( \xi ) }{ d \xi}+\frac{N c_{\vartheta_N}^2}{  N-1 }  \sum_\omega \bar{f}_{\omega}^2 ( \xi ) \frac{ V^2_\omega}{ \bar V^2} \,. \label{eq:SUN_matrix_Ddown}
\eeqn
\eeqs
The numerical analysis of the string stability relies on the eigenvalue of the following coupled Helmholtz equations
\beqn
&&  \hat \Oc \delta \Theta = \omega^2 \delta \Theta \,, 
\eeqn
where a negative eigenvalue of $\omega^2$ represents the unstable region.
The numerical code can be found in \cite{BianZYCode}.

\subsection{The stable regions}
\label{section:SUN_stable_region}

\para
Previously in the analysis of the non-topological strings~\cite{Eto:2024xvc,Bian:2026tco}, it was realized there are two major sources of the string instability. 
The first source is due to sufficiently strong magnetic fields of the $\Zc^\prime$ string.
With the string profile in Eq.~\eqref{eq:SUN_top_stringAnsatz}, the magnetic field couples to gauge bosons of $(\Wc_\mu^k \,, \Wc_\mu^{ k \, *} )$ with the strength of $\f{2 g_{\Zc^\prime} c_{ \vartheta_N }^2  }{N-1}$ according to Eqs.~\eqref{eq:SUN_matrix_Dup} and \eqref{eq:SUN_matrix_Ddown}.
The energy dispersion of the ${\rm U}(1)_{X^\prime}$-charged $(\Wc_\mu^k \,, \Wc_\mu^{ k \, *} )$ gauge bosons is given by the Landau levels of
\beqn
E^2 &=& ( 2 \ell + 1  - 2 S_z ) \f{2 g_{\Zc^\prime} c_{ \vartheta_N }^2  }{N-1} B_{\Zc^\prime } + p_z^2 + m_{ \Wc }^2 \,.
\eeqn
For the lowest Landau level of $\ell=0$ and $S_z=+1$, the energy of the ${\rm U}(1)_{X^\prime}$-charged $(\Wc_\mu^k \,, \Wc_\mu^{ k \, *} )$ gauge bosons will become negative when the magnetic field is stronger than 
\beqn
&& B_{ \Zc^\prime } > \f{ ( N-1) }{2 g_{ \Zc^\prime }  c_{\vartheta_N }^2 }  m_\Wc^2 =  \f{N }{4}  \f{ m_{ \Zc^\prime }^2}{ g_{\Zc^\prime } } \,,
\eeqn
with the gauge boson masses in Eqs.~\eqref{eq:Zp_masses} and \eqref{eq:offW_masses}.
The instability is due to the spin magnetic coupling term of $c_{ \vartheta_N }^2 \f{1}{ \xi} \f{\partial \bar \zeta (\xi) }{ \partial \xi}$ in the $\mathcal{D}_{ \uparrow}$ element.
Eq.~\eqref{eqs:SUN_topo_BCs_Infinity} gives $\bar\zeta(\infty)=-N n_1 c_{\tilde\beta}^2/g_{\Zc^\prime}$ for the $(n_1 \,,0)$ winding, and
$\bar\zeta(\infty)=-N n_1/g_{\Zc^\prime}$ for the $(n_1 \,, n_1)$ winding.
Thus, the former carries a smaller magnitude of the total $\Zc^\prime$ magnetic flux.
The effect on the spin-magnetic instability depends on the radial magnetic-field profile.

\para
The second source of the string instability is due to the heavy Higgs boson masses.
Equivalently, this is due to the large self couplings in the Higgs potential Eq.~\eqref{eq:SUN_twoHiggs_perturb_reduced}.
To highlight this effect, we take the semilocal limit of $\vartheta_N\to \f{\pi}{2}$, analogous to the $\vartheta_W\to \f{\pi}{2}$ limit for the embedded $Z$ string~\cite{Eto:2024xvc}.
The matrix elements in Eqs.~\eqref{eqs:SUN_matrix_stable} are reduced to
\beqs
\beqn
\mathcal{D}_{11}&\to& - \frac{ d^2}{ d \xi^2} - \frac{1}{\xi}\frac{ d }{ d \xi} +\frac{1}{\xi^2} \( n_1 + \ell+   \f{ g_{ \Zc^\prime} }{ N }  \bar{\zeta} ( \xi )  \)^2  \non
&& + \beta_1 c_{\tilde \beta}^2 ( \bar f_1^2 ( \xi) -1) + \beta_3 \(   c_{\tilde \beta}^2 ( \bar f_1^2 ( \xi) -1) +  s_{\tilde \beta}^2 ( \bar f_2^2 ( \xi) -1) \) + \beta_4 s_{\tilde \beta}^2 \bar f_2^2 ( \xi )  \,, \label{eq:SUN_semi_D11}\\[2mm]
\mathcal{D}_{22}&\to& - \frac{ d^2}{ d \xi^2} - \frac{1}{\xi}\frac{ d }{ d \xi}  +\frac{1}{\xi^2} \( n_2 + \ell+   \f{ g_{ \Zc^\prime} }{ N } \bar{\zeta} ( \xi )  \)^2   \non
&& + \beta_2 s_{\tilde \beta}^2 ( \bar f_2^2 ( \xi) -1) +   \beta_3 \(   c_{\tilde \beta}^2 ( \bar f_1^2 ( \xi) -1) +  s_{\tilde \beta}^2 ( \bar f_2^2 ( \xi) -1) \) + \beta_4 c_{\tilde \beta}^2 \bar f_1^2 ( \xi ) \,, \label{eq:SUN_semi_D22}\\[2mm]
\mathcal{D}_{12}&\to&  - \beta_4    \bar f_1( \xi ) \bar f_2 ( \xi )    s_{\tilde \beta} c_{\tilde \beta}  \,,\label{eq:SUN_semi_D12}\\[2mm]
\Bc_{ \uparrow/\downarrow \,, 1/2}&\to& 0\,, \\[2mm]
\mathcal{D}_{ \uparrow\,, -\ell} &\to & - \frac{ d^2}{ d \xi^2} - \frac{1}{\xi}\frac{ d }{ d \xi} +\frac{1}{\xi^2} \( \ell +1  \)^2  \,, \\[2mm]
\mathcal{D}_{ \downarrow \,, - \ell }& \to & - \frac{ d^2}{ d \xi^2} - \frac{1}{\xi}\frac{ d }{ d \xi} +\frac{1}{\xi^2} \( \ell - 1 \)^2 \,. 
\eeqn
\eeqs
By focusing on the diagonal $(\Dc_{11} \,, \Dc_{22} )$ terms in Eqs.~\eqref{eq:SUN_semi_D11} and \eqref{eq:SUN_semi_D22}, the larger inputs of $(\beta_1\,, \beta_2 \,, \beta_3)$ will destabilize the scalar perturbed modes, with the negative contributions from the $\bar f_{1\,, 2}^2 ( \xi) -1$ terms when one approaches to the string core.
Due to the Neumann boundary condition for the $(n_1\,, 0)$ $\Zc^\prime$ string in Eq.~\eqref{eq:SUN_topo_BCs_10}, the profile functions of $\bar f_2(\xi =0)$ are non-vanishing at the string core, as shown in the left panel of Fig.~\ref{fig:SUN_profiles}.
Therefore, an increasing and positive value of $\beta_4$ would enhance the stable region, as can be seen from Eq.~\eqref{eq:SUN_semi_D11}.
With the Dirichlet boundary condition for the $(n_1 \neq 0\,, n_2 \neq 0)$ winding configurations in Eq.~\eqref{eq:SUN_topo_BCs_11}, increasing the positive value of $\beta_4$ cannot enhance the stable region, which was the case for our previous analysis in Ref.~\cite{Bian:2026tco}.

\begin{figure}[htb]
\centering
\includegraphics[height=4.5cm]{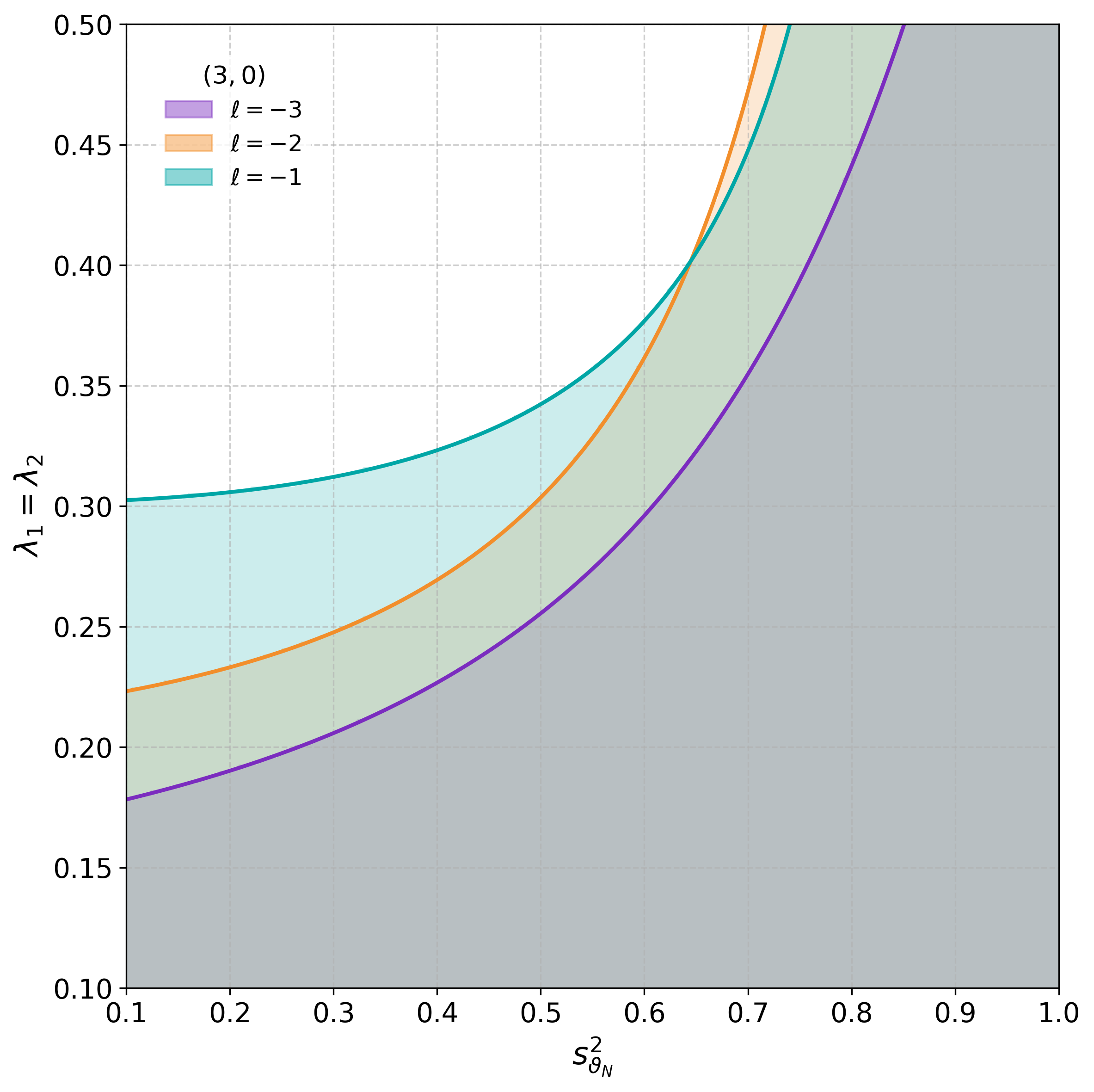}
\includegraphics[height=4.5cm]{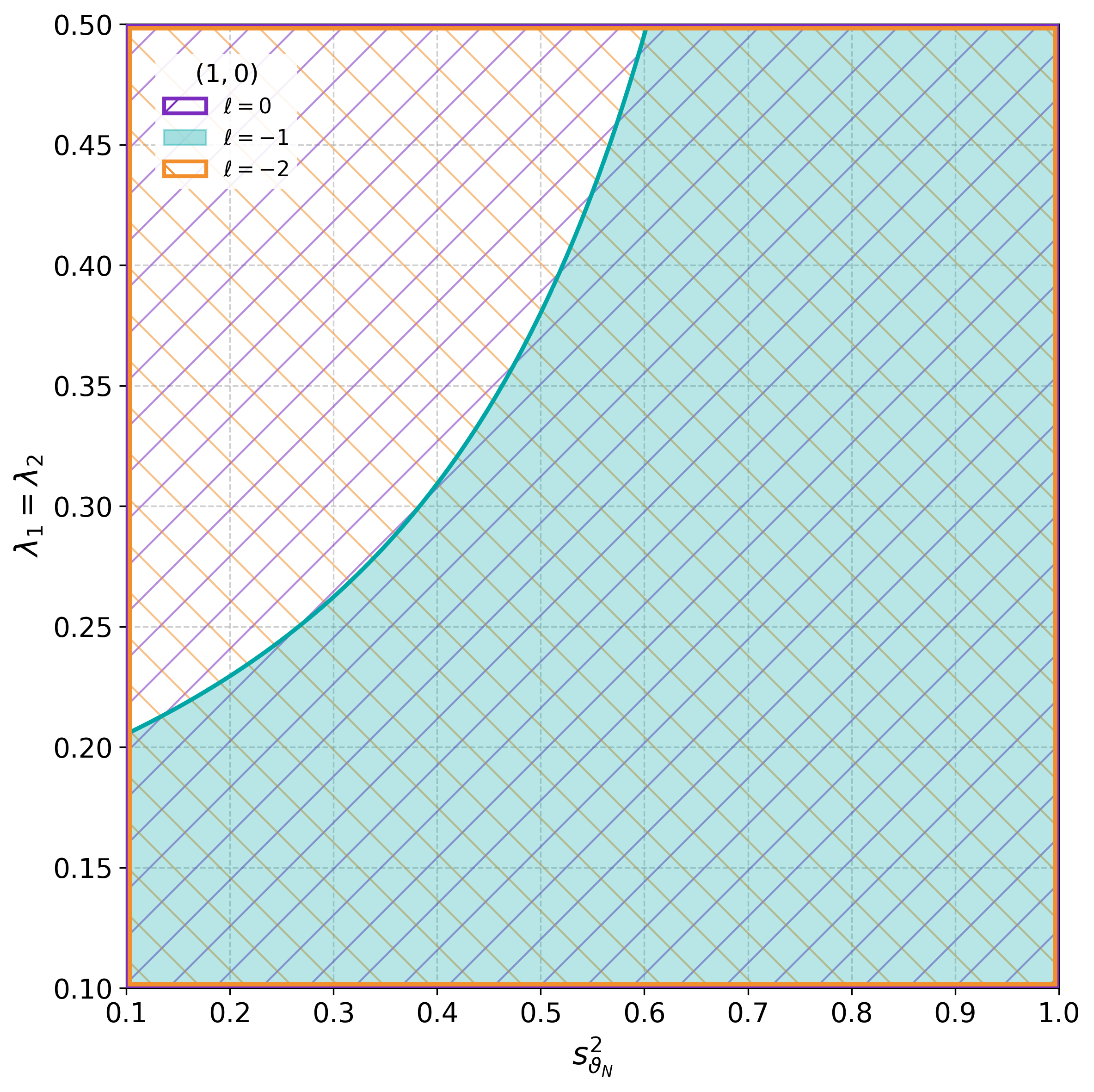}
\caption{Left panel: the shaded stable regions for the $(3\,,0)$ string with different Fourier modes of $\ell=(-3\,, -2\,, -1)$. 
Right panel: the shaded stable regions for the $(1\,,0)$ string with different Fourier modes of $\ell=(0\,, -1\,, -2)$.
Other parameters are $\lambda_1=\lambda_2$, $\lambda_3 = - \f{1}{2} \lambda_1 +0.005$, $\lambda_4=0.15$, $\tilde \beta=\f{\pi}{4}$, and $g_{\Zc^\prime}=1.06$ for the ${\rm SU}(3)\otimes {\rm U}(1)_X$ model.
}
\label{fig:SUN_Fourier_stabilities} 
\end{figure}

\begin{figure}[htb]
\centering
\includegraphics[height=4.2cm]{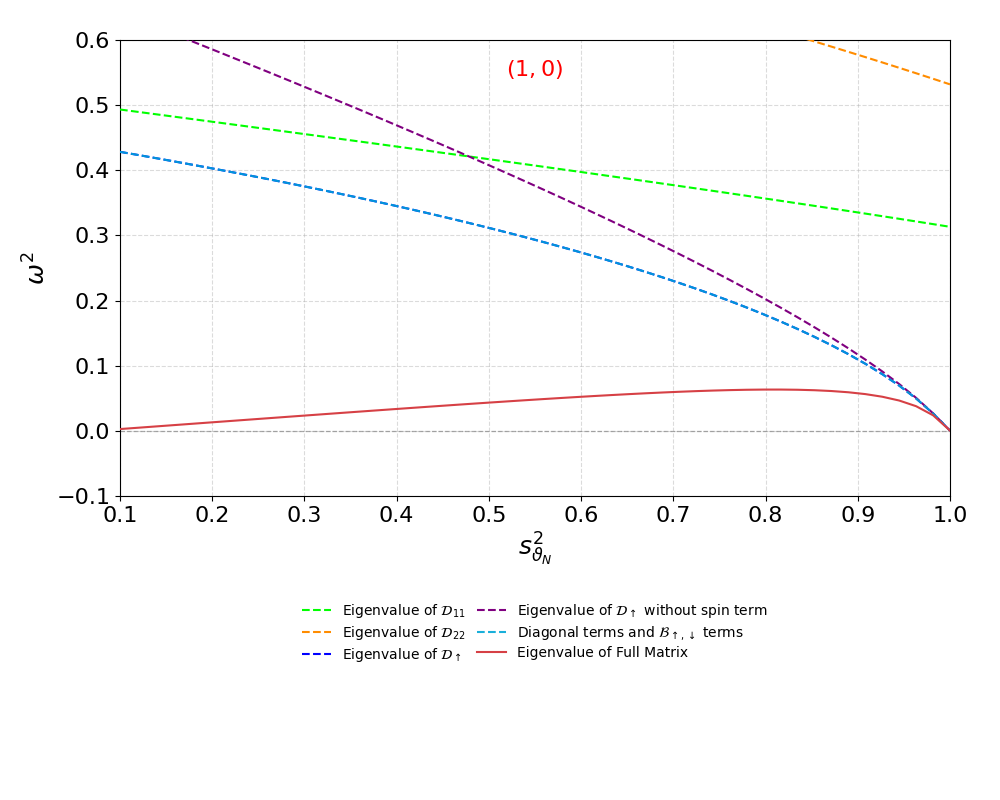}
\includegraphics[height=4.2cm]{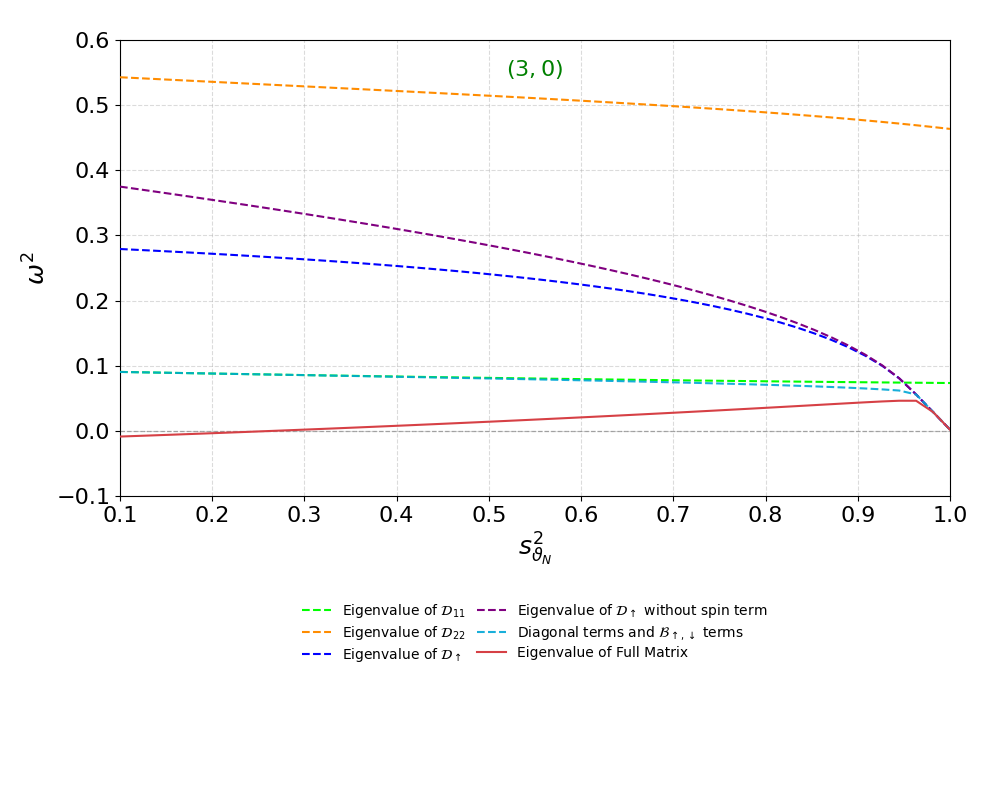}
\includegraphics[height=4.2cm]{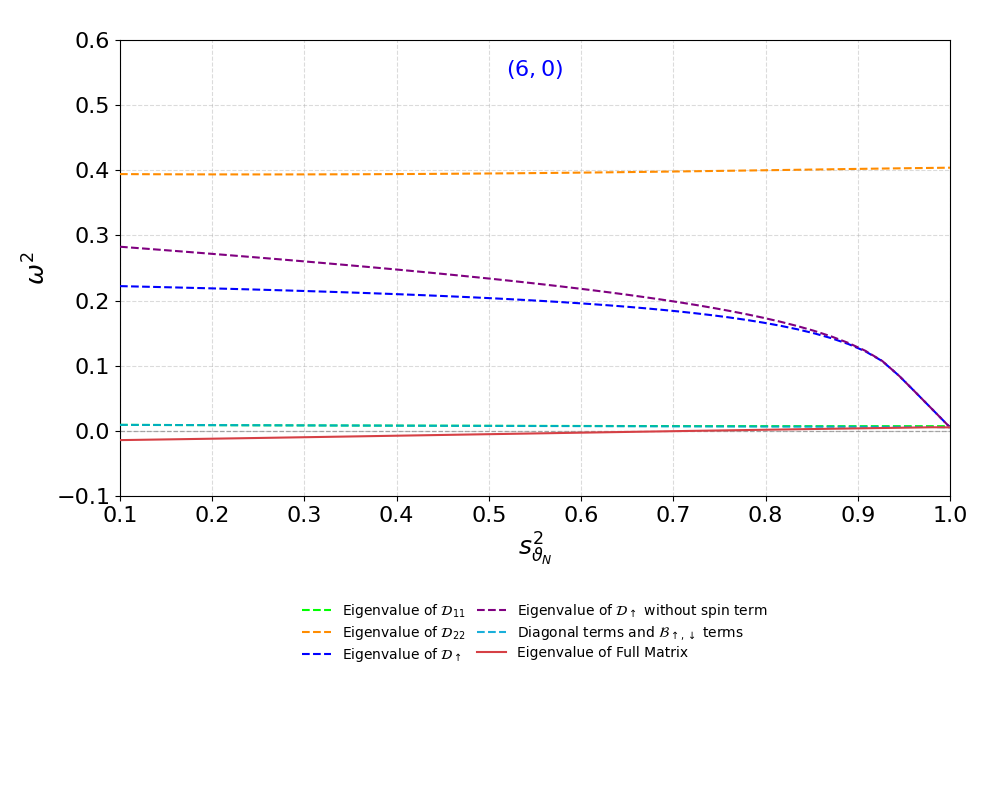}
\caption{The eigenvalues of $\omega^2$ of different stability matrix elements versus the mixing angle $s_{ \vartheta_N}^2$, for the $(1\,,0)$ string (left panel), the $(3\,,0)$ string (middle panel), and the $(6\,,0)$ string (right panel) configurations.
Other parameters are $\lambda_1=\lambda_2=0.2$, $\lambda_3 = -0.095$, $\lambda_4=0.15$, $\tilde \beta=\f{\pi}{4}$, and $g_{\Zc^\prime}=1.06$ for the ${\rm SU}(3)\otimes {\rm U}(1)_X$ model.
The Fourier modes are chosen to be $\ell=-n_1$.
}
\label{fig:SUN_omega2} 
\end{figure}

\para
Similar to the stability analysis of the electroweak string in Ref.~\cite{Goodband:1995he}, the stable regions should be obtained for different Fourier modes of the matrix elements in Eqs.~\eqref{eqs:SUN_matrix_stable}.
In Fig.~\ref{fig:SUN_Fourier_stabilities}, we display the shaded stable regions for the $(3\,,0)$ string (left panel) and the $(1\,,0)$ string (right panel), respectively.
From the shaded stable regions for the $(3\,,0)$ string, the most stringent constraints to the parameter regions are due to the $\ell=-3$ mode, as compared to the constraints from the $\ell=(-1\,,-2)$ modes.
From the shaded stable regions for the $(1\,,0)$ string, it turns out the parameter regions for $s_{\vartheta_N}^2 \in(0.1\,, 1.0)$ and $\lambda_1 = \lambda_2 \in(0.1\,, 0.5)$ for the $\ell=(0\,, -2)$ modes are completely stable, and they are represented by grids in order to highlight the constrained regions due to the $\ell=-1$ mode.
Turning to the stability matrix element of $\Dc_{11}$ in Eq.~\eqref{eq:SUN_semi_D11}, the choices of $\ell=-n_1$ suppress the positive contribution from the $\f{1}{\xi^2} \( n_1 + \ell + \f{g_{\Zc^\prime} }{N} \bar \zeta(\xi)  \)^2$ term most significantly.
Below, we will always take the Fourier mode of $\ell=-n_1$ for the specific $(n_1\,, 0)$ winding configuration.

\begin{figure}[htb]
\centering
\includegraphics[height=4.5cm]{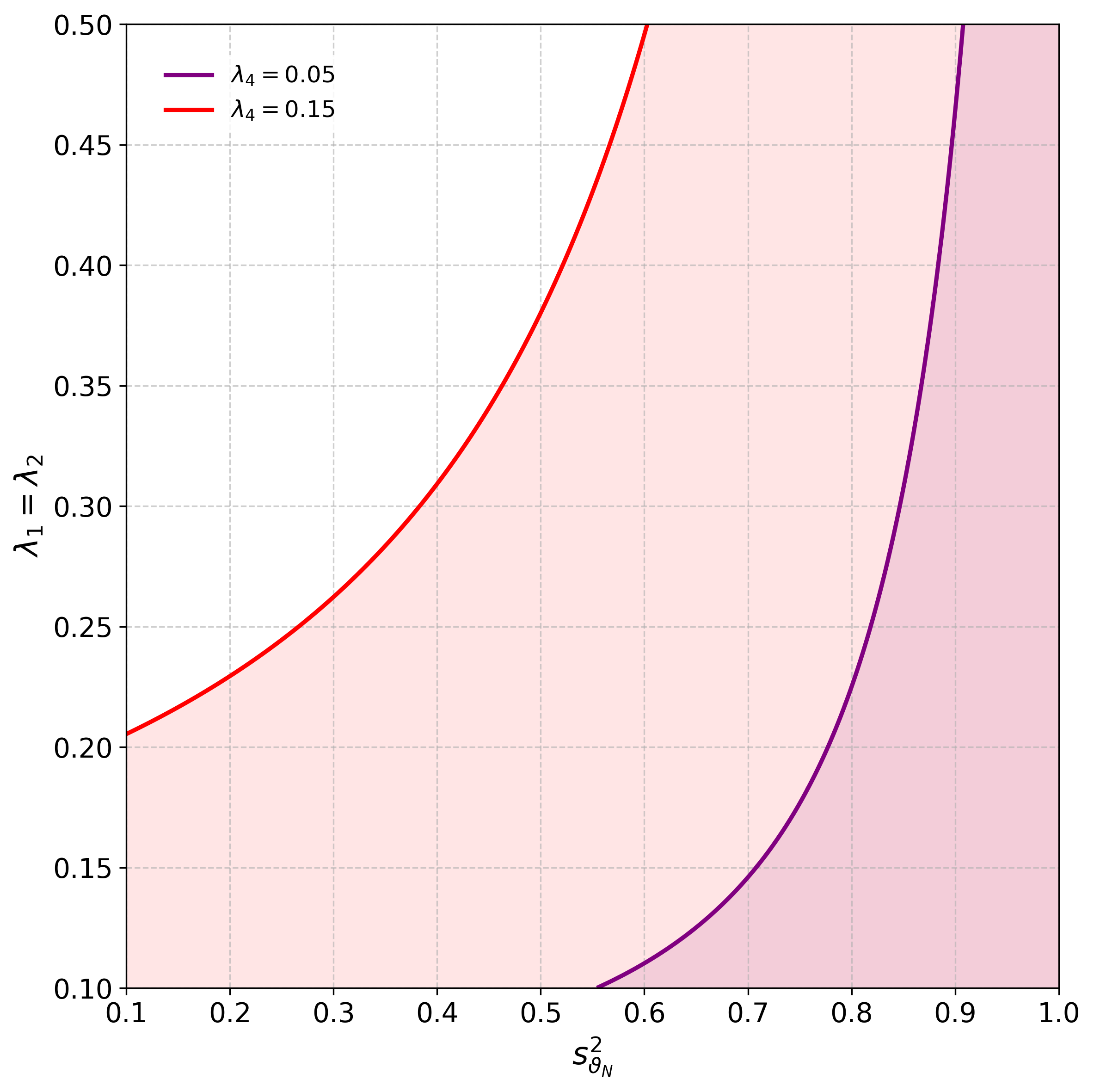}
\includegraphics[height=4.5cm]{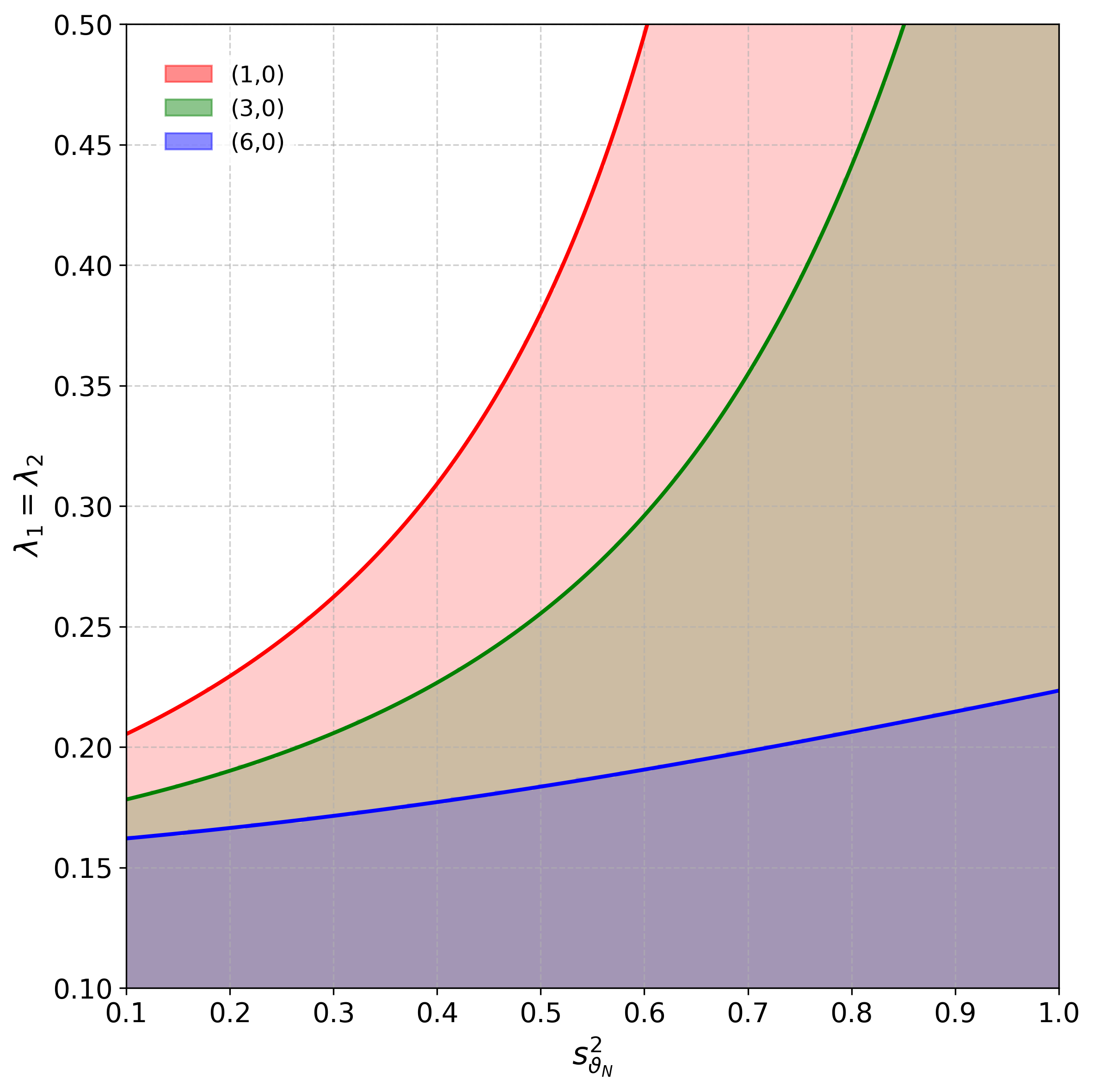}
\includegraphics[height=4.5cm]{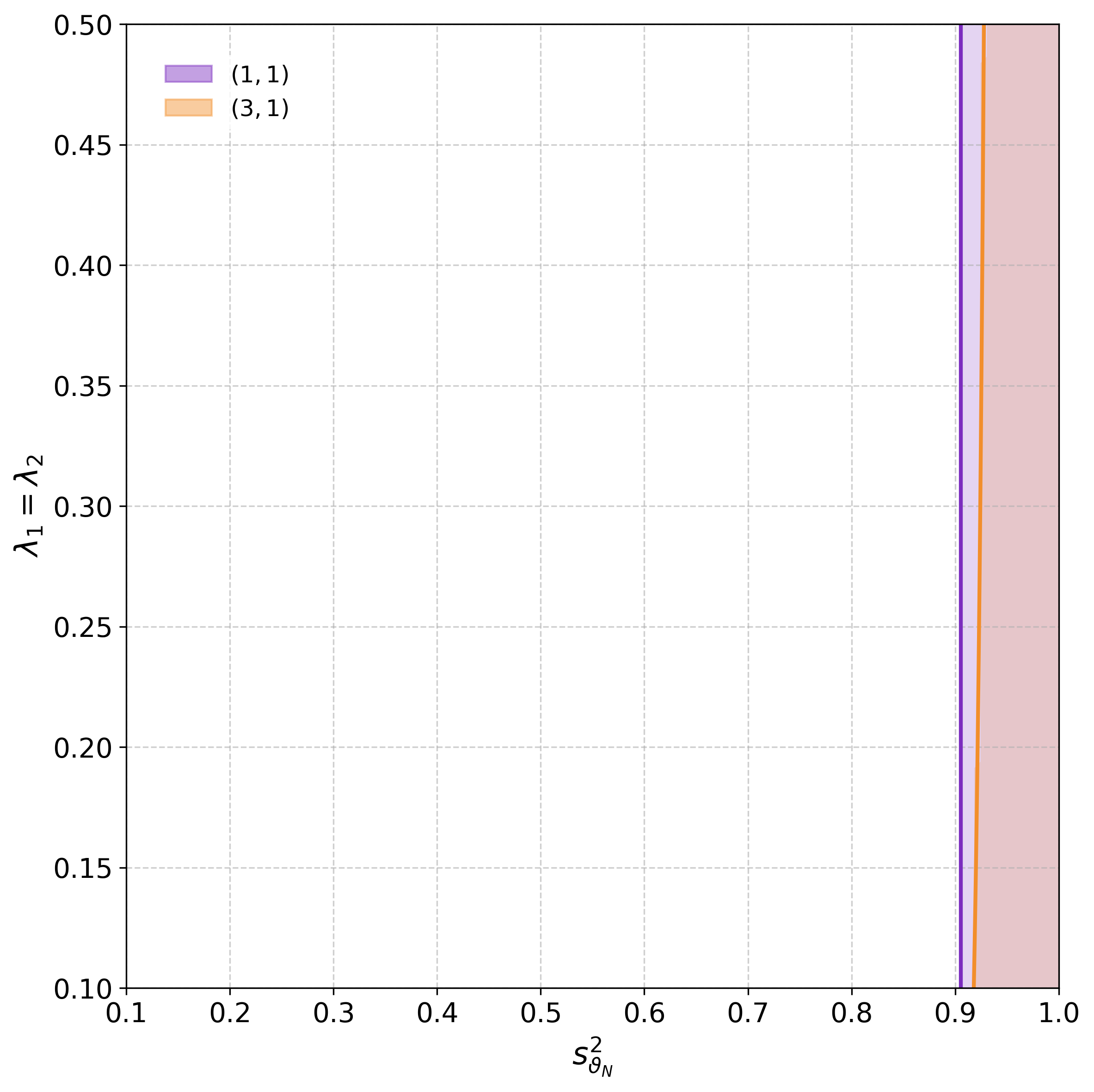}
\caption{Left panel: the shaded stable regions for the $(1\,,0)$ string with different inputs of $\lambda_4$. 
Middle panel: the shaded stable regions for different winding configurations of the $(1\,,0)$, $(3\,,0)$ and $(6\,,0)$ strings, with the fixed input of $\lambda_4=0.15$.
Right panel: the shaded stable regions for different winding configurations of the $(1\,,1)$ and $(3\,,1)$ strings, with the fixed input of $\lambda_4=0.15$.
Other parameters are $\lambda_1=\lambda_2$, $\lambda_3= -\f{1}{2} \lambda_1 +0.005$, $\tilde \beta=\f{\pi}{4}$ and $g_{\Zc^\prime}=1.06$, for the ${\rm SU}(3)\otimes {\rm U}(1)_X$ model.
}
\label{fig:SUN_stabilities} 
\end{figure}

\para
In Fig.~\ref{fig:SUN_omega2}, we display the eigenvalues of $\omega^2$ for different stability matrix elements in Eqs.~\eqref{eqs:SUN_matrix_stable}.
In all plots, the eigenvalues of the diagonal elements $( \Dc_{11} \,, \Dc_{22})$ are positive while decreasing as one varies $\vartheta_N$ from $0$ to $\f{\pi}{2}$.
The $\Dc_{\uparrow }$ elements contain the spin magnetic term, and the inclusion of this term always reduces the corresponding eigenvalues, as one compares the dashed purple curves and the dashed blue curves.
By further including the off-diagonal $\Bc_{\uparrow \,, \downarrow}$ elements, the joint effects of $\Dc_\uparrow$ and $\Bc_{\uparrow \,, \downarrow}$ further reduce the eigenvalues of $\omega^2$.
The ranges of $s_{\vartheta_N}^2$ for the positive $\omega^2$ match the constrained stable regions in the right panel of Fig.~\ref{fig:SUN_stabilities} for three different winding configurations.

\para
In Fig.~\ref{fig:SUN_stabilities}, we present the stable regions in the $( s_{\vartheta_N }^2 \,, \lambda_1=\lambda_2)$ plane for the topological strings of the ${\rm SU}(3) \otimes {\rm U}(1)_X$ model.
We always choose $\lambda_3= -\hf \lambda_1+0.005$, in order to obtain a positive contribution from the $\beta_3$ term in the $\Dc_{11}$ and $\Dc_{22}$ elements in Eqs.~\eqref{eq:SUN_stable_D11} and \eqref{eq:SUN_stable_D22}.
In the left panel, we display the shaded stable regions for two different inputs of $\lambda_4=0.05$ and $\lambda_4=0.15$, respectively, with the $(1\,,0)$ winding configuration.
Indeed, increasing of $\lambda_4$ significantly enhances the parameter ranges of $s_{\vartheta_N}^2$ for a stable string, as we have expected previously.
In the middle panel, we further display the shaded stable regions for the fixed input of $\lambda_4=0.15$, with different winding configurations of $(1\,,0)$, $(3\,,0)$ and $(6\,,0)$.
All three winding configurations allow the viable mixing angles of $0.1\lesssim s_{ \vartheta_N}^2 \lesssim 1.0$, together with the upper limits to the self couplings of $\lambda_{1\,,2}$.
For winding configurations with larger $n_1$, the stable regions are more significantly constrained, as compared to the $(1\,,0)$ winding configuration.
From the boundary conditions for the gauge fields in Eqs.~\eqref{eq:SUN_topo_BCs_10} and \eqref{eqs:SUN_topo_BCs_Infinity}, the larger winding numbers lead to larger spin magnetic coupling of $\f{1}{\xi} \f{d \bar \zeta(\xi) }{d \xi}$ in the matrix element $\Dc_\uparrow$.
Also from the left panel of Fig.~\ref{fig:SUN_profiles}, larger winding numbers of $n_1$ suppress the boundary values for the string profile function $\bar f_2(\xi)$ at the string core.
Correspondingly, the positive contributions with the positive $\lambda_4$ input in Eq.~\eqref{eq:SUN_stable_D11} are reduced, as compared to the $(1\,,0)$ string configuration.
Altogether, we expect a large $(n_1\,, 0)$ winding configuration can be stable only if the Higgs self couplings (or the physical masses of Higgs fields) are sufficiently small.
In the right panel, we exhibit the shaded stable regions for two $(1\,,1)$ and $(3\,,1)$ winding configurations for comparison. 
As we have mentioned, such configurations increase the asymptotic values of gauge fields outside of the string as compared to the $(1\,,0)$ and the $(3\,,0)$ winding configurations.
Also, due to the Dirichlet boundary conditions of $\bar f_{1\,,2}(\xi=0)=0$ in Eq.~\eqref{eq:SUN_topo_BCs_11}, a positive input of $\lambda_4$ can no longer enhance the stable region from the matrix elements.
Altogether, we find that the stable regions for the winding configurations of $(n_1\neq 0\,, n_2\neq 0)$ turn to the semilocal limit of $\vartheta_N\to \f{\pi}{2}$.
As we have previously discussed in Ref.~\cite{Bian:2026tco}, this limit is unrealistic in the extended gauge sectors based on the GUTs beyond the ${\rm SU}(5)$.

\vspace*{3mm}
\section{Conclusion}
\label{section:conclusion}

\para
In this paper, we have carried out the detailed studies of the topological $\Zc^\prime$ string in a class of ${\rm SU}(N) \otimes {\rm U}(1)_X$ models, for $N>2$ in general.
Two anti-fundamental Higgs fields are assumed to be responsible for the sequential symmetry breaking pattern.
The topological origin of the string is the global $\widetilde{\rm U}(1)$ symmetry that emerges in the Higgs potential \eqref{eq:SUN_twoHiggs_potential} when $\lambda_5=0$. 
This symmetry is spontaneously broken by the Higgs VEVs, giving rise to topologically stable string configurations.
With the most generic string profiles with arbitrary winding numbers of $(n_1\,, n_2)$ in Eq.~\eqref{eq:SUN_top_stringAnsatz}, we obtain the string profile functions, tensions, and the non-integer magnetic flux in Eq.~\eqref{eq:SUN_magFlux}.
For $n_1\neq n_2$, the asymptotic energy density does not vanish at $\xi=\infty$, so the string is a global one with divergent tension.
A finite tension is recovered only for the special case of $n_1=n_2$, which corresponds to the non-topological string configuration discussed previously.

\para
We then performed a detailed stability analysis by introducing time-dependent perturbations to both the Higgs and gauge fields around the $(n_1\,,n_2)$ string background.
After the Fourier expansion in the two-dimensional polar coordinates, the stability problem reduces to a set of coupled Helmholtz equations. 
Negative eigenvalues of the resulting stability matrix signal the unstable regions. 
For the $(n_1\,, 0)$ winding configuration, or equivalently the $(0\,, n_2)$ winding configuration, we found stable regions over a broad range of the mixing angle $\vartheta_N$ and the Higgs self couplings.
Larger winding numbers lead to more stringent constraints on the parameter space, while increasing the input of $\lambda_4$ enlarges the stable region.
For the general winding configuration of $(n_1 \neq 0\,, n_2\neq 0)$, stable regions exist only in the semilocal limit of $\vartheta_N\to \f{ \pi }{2}$.
Compared with the electroweak $Z$ string in the SM, which requires that $\vartheta_W \to \f{\pi}{2}$ and a light SM Higgs boson of $m_H < m_Z$, the topological $\Zc^\prime$ strings in the extended ${\rm SU}(N)\otimes {\rm U}(1)_X$ can be classically stable for a wider range of mixing angles and Higgs potential parameters, especially for single-winding configurations of $(n_1\,,0)$ and $(0\,,n_2)$.
This makes them potentially relevant for new physics scenarios with extended gauge symmetries and multiple Higgs fields.

\para
Several directions remain for future work. 
It would be interesting to include explicit $\lambda_5 \neq 0$ terms and study the resulting string-wall composites~\cite{Eto:2018hhg,Eto:2018tnk,Eto:2023orr}, as well as finite-temperature and cosmological evolution of these topological strings. 
Embedding the present analysis into realistic grand unified or the $\Gc_{331}$ models, and examining possible phenomenological signatures, would also be worthwhile. 
Finally, supersymmetric extensions~\cite{Davis:1997bs} and the role of fermionic zero modes~\cite{Earnshaw:1994jj,Liu:1995at,Davis:1999ec,Starkman:2000bq,Starkman:2001tc,Eto:2024xvc,Harigaya:2024hah} on these strings could provide further insight into their stability and observational consequences.

\vspace*{3mm}
\section*{Acknowledgments}
%
%
\para
We would like to thank Maolin Zhou for very useful discussions. 
N.C. thanks Shandong University and Nanjing Normal University for hospitality when preparing this work.
This work is partially supported by the National Natural Science Foundation of China (under Grant No. 12275140) and Nankai University.

\providecommand{\href}[2]{#2}\begingroup\raggedright\endgroup

\end{document}